 \documentclass[prb,aps,twocolumn,showpacs]{revtex4} 
\usepackage{graphicx} 
\usepackage{tabularx}
\usepackage{dcolumn} 
\usepackage{color}
\usepackage{bm}
\usepackage{amsmath}
\usepackage{amssymb}
\usepackage{newtxtext, newtxmath}

\begin{document} 
 
\title{
Theoretical studies for identifying horizontal line nodes via angle-resolved density of states measurements\\
---Application to Sr$_2$RuO$_{4}$---
}


\author{Kazushige Machida$^{\ast}$, Koki Irie$^{\ast}$, Katsuhiro Suzuki$^{\dagger}$, Hiroaki Ikeda$^{\ast}$} 
\affiliation{$^{\ast}$Department of Physics and $^{\dagger}$Research Organization of Science and Technology, 
Ritsumeikan University, 
Kusatsu 525-8577, Japan} 
\author{Yasumasa Tsutsumi}  
\affiliation{
Department of Basic Science, University of Tokyo, Meguro, Tokyo 153-8902, Japan}
\affiliation{
RIKEN Center for Emergent Matter Science, Wako, Saitama 351-0198, Japan}

\date{\today}

\begin{abstract}
On the basis of the microscopic quasi-classical Eilenberger theory, we analyze the recent angle-resolved 
specific heat experiment carried out at low temperature for Sr$_2$RuO$_{4}$ to identify the superconducting gap symmetry,
comprising either horizontal or vertical line nodes relative to the tetragonal crystal symmetry. 
Several characteristics, in particular, the landscape of the in-plane oscillation amplitude
$A_4(B, T)$ with a definite sign for almost the entire $B$-$T$ plane are best explained by the horizontal line node symmetry,
 especially when the multiband effect and Pauli paramagnetic effect are taken into account.
The present analysis of $A_4(B,T)$ with definite sign points to the presence of an anomalous field region
at a lower temperature in the experimental data, whose origin is investigated. 
Our theory demonstrates the application and uniqueness of the field-rotating 
thermodynamic measurements in uncovering the precise gap structure for target materials.
\end{abstract}

 
 
\maketitle 

\section{Introduction}
Sr$_2$RuO$_4$~\cite{maeno0} is a prime candidate of a chiral $p$-wave superconductor\cite{mackenzie,maeno}.
Although many experimental and theoretical studies have been devoted to identifying its pairing symmetry,
which involves its spin structure, i.e., spin triplet or spin singlet and its 
orbital or energy gap structure in $k$-space.
Both its spin and orbital structures remain elusive and controversial\cite{maeno,mackenzie2}.
The early nuclear magnetic resonance (NMR) experiment by Ishida \textit{et al.}\cite{ishida} detected no
change of the Knight shift below $T_c\sim 1.5$K for field direction parallel to the $ab$ plane, thus
leading to the naive interpretation that the spin structure is triplet where the $d$-vector lies parallel to the $c$-axis.
However, later experiments\cite{murakawa1,murakawa2} for $H\parallel c$ unexpectedly detected no
change at all. Therefore, the naive interpretation did not hold anymore.
Such results must be regarded with caution.
It is difficult to be convinced of $d$-vector rotation under an applied field as low as 300 gauss.
Kim \textit{et al.} \cite{kim} estimated the strength of the spin-orbit coupling to lock the $d$-vector to the lattice
and concluded that the $d$-vector rotation interpretation is not correct.
Simultaneously they
proposed that the spin structure is spin singlet in this system.
 ``Decisive'' experiments\cite{nelson,jang,yasui} that claim spin-triplet pairing
in this system must be carefully scrutinized.
 Among them the observation\cite{jang} of a half-quantum fluxoid is definitive evidence because it is only 
realized for spin-triplet pairing. 

In a recent series of bulk thermodynamic measurements of the magnetocaloric effect\cite{kajikawa},
specific heat\cite{yonezawa}, and magnetization\cite{kittaka} for $H\parallel ab$ all detected a first order transition at $H_{c2}$ 
at low temperatures $T<$0.8 K.
By estimating the entropy and magnetization jumps at the first order transition, it was concluded that 
the quasi-particle density of states (DOS) decreases below $T_c$ upon entering the superconducting state.
This means that the spin susceptibility decreases in the superconducting state, thus completely
contradicting the Knight shift experiments\cite{ishida,murakawa1,murakawa2}.
The bulk measurements~\cite{kajikawa,yonezawa,kittaka} clearly point to a typical spin-singlet superconductor with strong Pauli paramagnetic 
effect (PPE).

This picture is also supported by neutron scattering experiments\cite{morten1,morten2} and corresponding 
theoretical analyses\cite{amano1,amano2,nakaiFF}, which find an anisotropic 
triangular vortex lattice with anisotropy $\Gamma_{VL}\sim$60 for $H\parallel ab$.
When compared with the upper critical field anisotropy of $\Gamma_{H_{c2}}=H_{c2}^{ab}/H_{c2}^c\sim20$,
it appears that the in-plane $H_{c2}^{ab}$ is strongly suppressed by PPE. The intrinsic
orbital anisotropy is at least 60, which nicely coincides with the Fermi velocity anisotropy $\Gamma_{\beta}=60$
for the $\beta$-band observed by dHvA experiments\cite{mackenzie,bergemann}.

As for the orbital symmetry of the pairing function or the gap structure, discussion 
and debate~\cite{mackenzie,maeno,mackenzie2}
continue. 
Since the existence of linear line nodes has already been ascertained 
by a variety of thermodynamic measurements\cite{mackenzie,maeno}, such as
specific heat, ultrasound attenuation, and thermal conductivity etc, the remaining questions are

(1) Where are the linear line nodes, whether vertical or horizontal line nodes relative to the $ab$-plane?

(2) Which band is responsible for them among the three bands, $\alpha$-, $\beta$- and $\gamma$-band 
or are they all responsible?

(3) Is the gap structure symmetry protected or band-dependent?

Angle-resolved thermodynamic measurements are now recognized as a 
quite powerful technique that can detect the nodal position in $k$-space~\cite{miranovic1,miranovic2,sakaki1,sakaki2}.
Deguchi \textit{et al.}~\cite{deguchi1,deguchi2}
carried out a pioneering angle-resolved specific heat experiment on Sr$_2$RuO$_4$ and find
four-fold oscillation with the (100) minimum parallel to the $a$-axis when rotating the $B$ field in the
$ab$ plane. In their interpretation of their results, the (100) direction is the nodal direction, thus suggesting a d$_{xy}$-like
gap structure. However, subsequent theoretical studies~\cite{vekhter,hiragi}
have shown that if this is true,
oscillation pattern reversal or sign changing temperature at $T_{ch}\simeq0.15T_c$ must occur.
Unfortunately, Deguchi \textit{et al.}'s measurement barely reached this temperature region.
Therefore, the d$_{xy}$-like gap structure has not been confirmed.

Recently Hassinger \textit{et al.}~\cite{hassinger} claimed the presence of vertical line nodes on all bands 
based on their analysis of thermalconductivity 
data taken at low $T$, whereas, in a recent neutron scattering experiment~\cite{braden}, there was an
absence of the 
expected spin resonance at $Q=(0.3, 0.3, 0)$ [in reciprocal lattice units] in $k$-space 
associated with the vertical line nodes; thus the results are incompatible with Hassinger \textit{et al.}'s claim.

Here, we study the gap structure problem, for either horizontal line nodes (HLN) or
vertical line nodes (VLN) by analyzing the recent angle-resolved specific heat data
at lower temperatures down to 60 mK ($=0.04T_c)$~\cite{kittaka0}. The experimental results
are summarized as follows:

(I) The expected sign change of the oscillation amplitude A$_4(B, T)$
at $T_{ch}\simeq0.15T_c$ and $B_{ch}\simeq0.3B_{c2}$ for VLN in the single band case (see Figs.11 -- 13 in [\onlinecite{hiragi}])
is absent down to 60mK ($=0.04T_c)$ up to $B_{c2}$.

(I\hspace{-.1em}I) A$_4(B)$ tends to decrease toward higher fields after texhibiting a broad maximum
as $B$ is increased at lower $T$ (see Fig.~\ref{fig4-2-2}).

(I\hspace{-.1em}I\hspace{-.1em}I) A$_4(T)$ monotonically decreases upon increasing $T$ and tends to vanish
around $T\simeq0.2$ -- $0.3T_c$, which is quite low compared with the typical VLN case\cite{hiragi}
where A$_4(T)$ persists at least up to $T\simeq0.4$ -- $0.5T_c$ after exhibiting the sign change.

(I\hspace{-.1em}V) A$_4(B, T)$ shows A$_4(B, T)>0$ as functions of both $B$ and $T$, namely the (100) direction is
always specific heat minimum except just below $B_{c2}$ at low $T$.
This landscape of A$_4(B, T)$ differs substantially from that of VLN\cite{hiragi}
 where a local maximum, local minimum, 
and the sign changing line in the $B$-$T$ plane (see Fig.~\ref{fig2-13-2}(b))
are all present.

We investigate the origin of such characteristics via a microscopic quasi-classical Eilenberger framework~\cite{eilenberger}
valid for $k_F\xi\gg 1$ ($k_F$ the Fermi wave number and $\xi$ the coherence length), which is well met
for Sr$_2$RuO$_4$, to identify the gap structure of the Sr$_2$RuO$_4$ system.
Simultaneously, we investigate the validity and limitations of the semiclassical concept of the Doppler shift\cite{volovik} which
is conveniently applied to the oscillation phenomena\cite{vekhter}.
Needless to say, the Doppler shift itself is a a universally correct, fundamental physical concept with
wide applications. We find this semiclassical picture based on the Doppler shift applied to the quasi-particles in 
the vortex state, which we call the Doppler shift picture,
to be quite useful in understanding the thermodynamic oscillation
phenomena in a superconductor. However, some care is required when applying it to an actual situation.

This paper is organized as follows:
first, we introduce the formulation based on the microscopic quasi-classical 
Eilenberger theory as well as its approximate solution of the Kramer-Pesch approximation (KPA).
The modeling of the Fermi surfaces for our target material Sr$_2$RuO$_4$ is also introduced in Sec.~I\hspace{-.1em}I.
Then we examine the angle-resolved density of states in order to analyze the angle-resolved specific heat data~\cite{kittaka0} 
for Sr$_2$RuO$_4$ when the gap structure has the horizontal line nodes (HLN).
The calculations are done both for the full self-consistent solution of the 
Eileberger equation and for the KPA solutions. We also take into account the Pauli paramagnetic effect (PPE)
for the full solutions. The landscape of the DOS oscillation amplitude A$_4(B, T)$ is constructed
without and with PPE in Sec.~I\hspace{-.1em}I\hspace{-.1em}I. In the next Sec.~I\hspace{-.1em}V we examine the vertical line nodes (VLN) case comparatively.
Here the multiband effect, which crucially influences the specific heat oscillations, is discussed in detail.
In Sec.~V we analyze the experimental data on the specific heat oscillation~\cite{kittaka0}, at which point we emphasize that
the HLN scenario is far superior to the VLN one; we also show that our analysis reveals the presence of an 
anomalous high field region just below $B_{c2}$, which may be the first evidence for the FFLO
expected for this super-clean system.
Finally, we summarize the overall picture for the pairing symmetry in Sr$_2$RuO$_4$
and share future prospects of the material.

We note here that our earlier work~\cite{hiragi} thoroughly discusses the VLN case
by solving the full Eilenberger equation for the same quasi 2D cylindrical model with 
and without PPE and constructs the A$_4(B, T)$ landscapes.
The present paper should be regarded as an extension to the HLN case.

\section{Formulation and modeling}
\subsection{Eilenberger equation}
Quasiclassical Green's functions 
$f(\omega_n , {\bf p},{\bf r})$, 
$f^\dagger(\omega_n , {\bf p},{\bf r})$, and
$g(\omega_n , {\bf p},{\bf r})$ depend on the direction of the Fermi momentum $\bm{p}$, 
the center-of-mass coordinate $\bm{r}$ for the Cooper pair, and Matsubara frequency 
$\omega_n\!=\!(2n\!+\!1)\pi T$ with $n\!\in\!\mathbb{Z}$.
They are calculated in a unit cell of the triangle vortex lattice by solving the 
Eilenberger equation~\cite{eilenberger} for clean type II superconductors as follows:
\begin{eqnarray} && 
\left\{ \omega_n + {\rm i}{\mu} B({\bf r}) 
        + {\bf v}_{\rm F} \cdot 
           \left(\nabla + {\rm i}{\bf A}({\bf r}) \right) \right\} f 
= \Delta({\bf r}) g, 
\nonumber \\ && 
\left\{ \omega_n + {\rm i}{\mu} B({\bf r}) 
        - {\bf v}_{\rm F} \cdot 
           \left(\nabla - {\rm i}{\bf A}({\bf r}) \right) \right\} f^\dagger 
= \Delta^\ast({\bf r}) g , 
\label{eq:Eil} 
\end{eqnarray} 
with 
\begin{eqnarray} && 
{\bf v}_{\rm F} \cdot {\nabla} g 
= \Delta^\ast({\bf r}) f 
- \Delta({\bf r}) f^\dagger,
\end{eqnarray}
where the normalization 
$g=(1-ff^\dagger)^{1/2}$ is imposed. We take into account the Pauli paramagnetic effect
through the Maki parameter ${\mu}=\mu_{\rm B} B_0/\pi T_{\rm c}$. 
The Fermi velocity is ${\bf v}_{\rm F}$.
We scale length, temperature, and the magnetic field
in units of $\xi_0$, $T_c$, and 
$B_0$, respectively, 
where $\xi_0=\hbar v_{{\rm F}}/2\pi T_{\rm c}$ and 
$B_0=\phi_0 /2 \pi \xi_0^2$ ($k_{\rm B}=1$).
The vector potential 
${\bf A}=\frac{1}{2}\bar{{\bf B}}\times{\bf r}+{\bf a}({\bf r})$
is related to the internal field as 
${\bf B}({\bf r})=\nabla\times {\bf A}
 =(B_x({\bf r}),B_y({\bf r}),B_z({\bf r}))$ 
with $\bar{\bf B}=(0,0,\bar{B})$, 
$B_z({\bf r})=\bar{B}+b_z({\bf r})$ and 
$(B_x,B_y,b_z)=\nabla\times {\bf a}$. 

The pairing potential $\Delta({\bf r})$ 
is calculated by the gap equation 
\begin{eqnarray}
\Delta({\bf r})
= \pi g_0N_0 T \sum_{0 \le \omega_n \le \omega_{\rm cut}} 
\left\langle 
    f +{f^\dagger}^\ast 
\right\rangle_{{\bf p}} 
\label{eq:scD}
\end{eqnarray}
where 
$g_0$ is the pairing interaction and 
$N_0$ the density of states at the Fermi energy in the normal state. 
$g_0$ is defined by the cutoff energy $\omega_{\rm c}$ as 
$(g_0N_0)^{-1} = \ln T+2\,T\sum_{\omega_{n>0}}^{\omega_{\rm c}}\,\omega_n^{-1}$.
We carry out calculations using the cutoff $\omega_{\rm c}=20 T_{\rm c}$. 
The current equation used to obtain ${\bf a}({\bf r})$ is given by 
\begin{eqnarray}
\nabla\times \nabla \times {\bf a}({\bf r}) 
={\bf j}_{\rm s}({\bf r})+\nabla\times {\bf M}_{\rm para}({\bf r})
\label{eq:rotA}
\end{eqnarray} 
where the screening current is
\begin{eqnarray} && 
{\bf j}_{\rm s}({\bf r})
=-\frac{2T}{{{\kappa}}^2} \sum_{0 \le \omega_n} 
 \left\langle {\bf v}_{\rm F} 
         {\rm Im}\{ g \} 
 \right\rangle_{{\bf p}}, 
\label{eq:scH} 
\end{eqnarray} 
and the paramagnetic moment is given by
\begin{eqnarray} && 
M_{\rm para}({\bf r})
=M_0 \left( 
\frac{B({\bf r})}{\bar{B}} 
- \frac{2T}{{\mu} \bar{B} } 
\sum_{0 \le \omega_n} \left\langle {\rm Im} \left\{ g \right\} 
 \right\rangle_{{\bf p}}
\right).
\label{eq:scM} 
\end{eqnarray} 
Here, the normal state paramagnetic moment 
$M_0 = ({{\mu}}/{{\kappa}})^2 \bar{B} $, and 
${\kappa}=B_0/\pi T_{\rm c}\sqrt{8\pi N_0}$.  
The Ginzburg-Landau (GL) parameter $\kappa$ 
is the ratio of the penetration depth to the coherence length for 
$\bar{\bf B}\parallel c$. 

We set the unit vectors of the vortex lattice as
${\bf u}_1=c({\alpha}/{2},-{\sqrt{3}}/{2}), {\bf u}_2=c({\alpha}/{2}, {\sqrt{3}}/{2})$
with $c^2=2 \phi_0/ (\sqrt{3} \alpha \bar{B})$ and 
$\alpha=3 \Gamma(\theta)$~\cite{hiragi}.
Thus $\Gamma(\theta)$ expresses the anisotropy of the system
through the deformation of the hexagonal vortex unit cell in terms of $\alpha$.
$\phi_0$ is the flux quantum, and $\bar{B}$ is the average flux density. 
By solving the above equations iteratively,
we obtain self-consistent solutions of 
$\Delta({\bf r})$, ${\bf A}({\bf r})$, and 
the quasiclassical Green's functions~\cite{ichioka1,ichioka2,ichioka3}. 
We calculate the electronic state by knowing the quasiclassical Green function 
$g(\bm{p},\bm{r},\omega_n)$ where $i\omega_n\!\rightarrow\! E\!+\!i\eta$.
The density of states (DOS) is given by
\begin{align}
N(E)\!=\!N_0\!\left\langle{\rm Re}\left[g(\bm{p},\bm{r},\omega_n)|_{i\omega_n\!\rightarrow\! E\!+\!i\eta }\right]\right\rangle_{\bm{r,p}}\!,
\end{align}
where $\langle\cdots\rangle_{\bm r,\bm p}$ indicates the spatial average over a vortex unit cell and momentum average
over the Fermi surface.

\subsection{Kramer--Pesch approximation (KPA)}
One can obtain an approximate solution of Eq.~\eqref{eq:Eil} 
within Kramer-Pesch approximation (KPA)~\cite{Nagai2006,Nagai2011} without resorting to
heavy numerical computations when solving the full self-consistent Eilenberger equation.
A one-vortex solution of Eq.~\eqref{eq:Eil} valid for the low energy regime $E\sim 0$ is given by~\cite{Nagai2011}
\begin{align}
\frac{N(\bm{r},E=0)}{N_0}=\left\langle\frac{v_{\perp}(\bm{p})e^{-u(s)}}{C(y,\bm{p})}\frac{\eta}{E^2(y,\bm{p})+\eta^2}\right\rangle_{\bm{p}}
\end{align}
with
\begin{align}
u(s)=2{|d({\bm p})|\over v_{\perp}({\bm p})}\int^s_0\Delta_{\infty}f(s',y){s'\over\sqrt{s'^2+y^2}}ds'
\end{align}
where $d({\bm p})$ is the angle dependence of the gap function, while $\Delta_{\infty}$ is the order parameter far
from vortex core.
$\bm{v}_{\perp}(\bm{p})$ is a projection of $\bm{v}(\bm{p})$ into the $ab$ plane and $(s,y)$ 
is a coordinate of the plane with respect to the angle of $\bm{v}_{\perp}(\bm{p})$. 
We parameterize $\Delta(\bm{r})=f(s,y)e^{i\phi}$, then $C(y,\bm{p})$, and $E(y,\bm{p})$ 
are expressed by $f(s,y)$ and given as follows:
\begin{align}
f(s,y)={\sqrt{s^2+y^2}\over \sqrt{s^2+y^2+\xi_{0\perp}^2}},\\
C(y,{\bm p})=2 \sqrt{y^2+\xi_{0\perp}^2}K_1(r_0(y,{\bm p})),\\
E(y,{\bm p})=|d({\bm p})|\Delta_{\infty}{K_0(r_0(y,{\bm p}))\over K_1(r_0(y,{\bm p}))}
{y \over \sqrt{y^2+\xi_{0\perp}^2}},\\
r_0(y,{\bm p})=2{|d({\bm p})|\over v_{\perp}({\bm p})}\Delta_{\infty}\sqrt{y^2+\xi_{0\perp}^2}.
\end{align}
Here, $K_0(r_0(y,{\bm p}))$ and $K_1(r_0(y,{\bm p}))$ are modified Bessel functions.

Within this one-vortex approximation, one cannot consider a vortex lattice formation.
The magnetic field effect appears as an integral radius of $\!\langle N(\bm{r},E=0)\rangle_{\bm{r}}$,
namely
\begin{align}
\!\langle N(\bm{r},E=0)\rangle_{\bm{r}}={1\over \pi r_a^2}\int^{r_a}_0drN(r,E=0).
\end{align}
Here we assume a circular Wigner-Seitz cell for each vortex whose radius $r_a$ is given by 
$r_a/\xi_{0\perp}=\sqrt {B_{c2}/B}$, that is, at $B=B_{c2}=\phi_0/\pi\xi_{0\perp}^2$ vortices touch 
each other with the coherence length $\xi_{0\perp}$.
In the KPA calculations we deal with the effect of the Fermi velocity anisotropy 
within the change of the coherence length along the $ab$ plane.
We confirm that the KPA results qualitatively coincide with those from the full Eilenberger solution.

\subsection{Modeling of the Fermi surfaces}
As a model of the Fermi surface, we use a quasi-two-dimensional 
Fermi surface with a rippled cylinder shape. 
The Fermi velocity is assumed to be 
${\bf v}_F=(v_a,v_b,v_c)\propto(v_a(\phi),v_b(\phi),\tilde{v}_z \sin p_c)$ 
at 
${\bf p}=(p_a,p_b,p_c)\propto(p_{\rm F}\cos\phi, p_{\rm F}\sin\phi,p_c)$
on the Fermi surface which we also used in our previous work~\cite{hiragi}. 
We consider a case $\tilde{v}_z=1/\Gamma$,
to produce a large anisotropy ratio of the coherence lengths
of the in-plane $ \xi_{ab}$ and out-of plane $\xi_{c}$.
The vortex lattice anisotropy $\Gamma_{VL}$, which was observed to be $\sim60$,
is determined via the free energy minimum
after solving the Eilenberger equation and depends on the gap structure 
and on the presence or absence of PPE (see Refs. \onlinecite{amano1} and \onlinecite{nakaiFF} for detail).

The magnetic field orientation is tilted by $\theta$ from the $c$ axis 
toward the $ab$ plane. 
We use the following formula for the general anisotropic ratio $\Gamma(\theta)$ as
\begin{eqnarray} && 
\Gamma(\theta) ={1\over{\sqrt{\cos^2\theta+\Gamma^{-2}\sin^2\theta}}}.
\end{eqnarray} 

Considering the material of study, ${\rm Sr_2RuO_4}$~\cite{mackenzie,maeno},
we choose $\kappa=2.7$ and the anisotropy ratio 
$\Gamma(\theta=90^{\circ})\equiv\Gamma=60$.
We note that the $\Gamma$ value does not significantly influence
the following in-plane oscillation calculations.

Since we set the $z$-axis to the vortex line direction, 
the coordinate ${\bf r}=(x,y,z)$ for the vortex structure is related to the 
crystal coordinate $(a,b,c)$ 
as $(x,y,z)=(a,b \cos\theta + c \sin\theta,c \cos\theta -b \sin\theta)$. 

\begin{figure}[tbp]
\includegraphics[width=8cm]{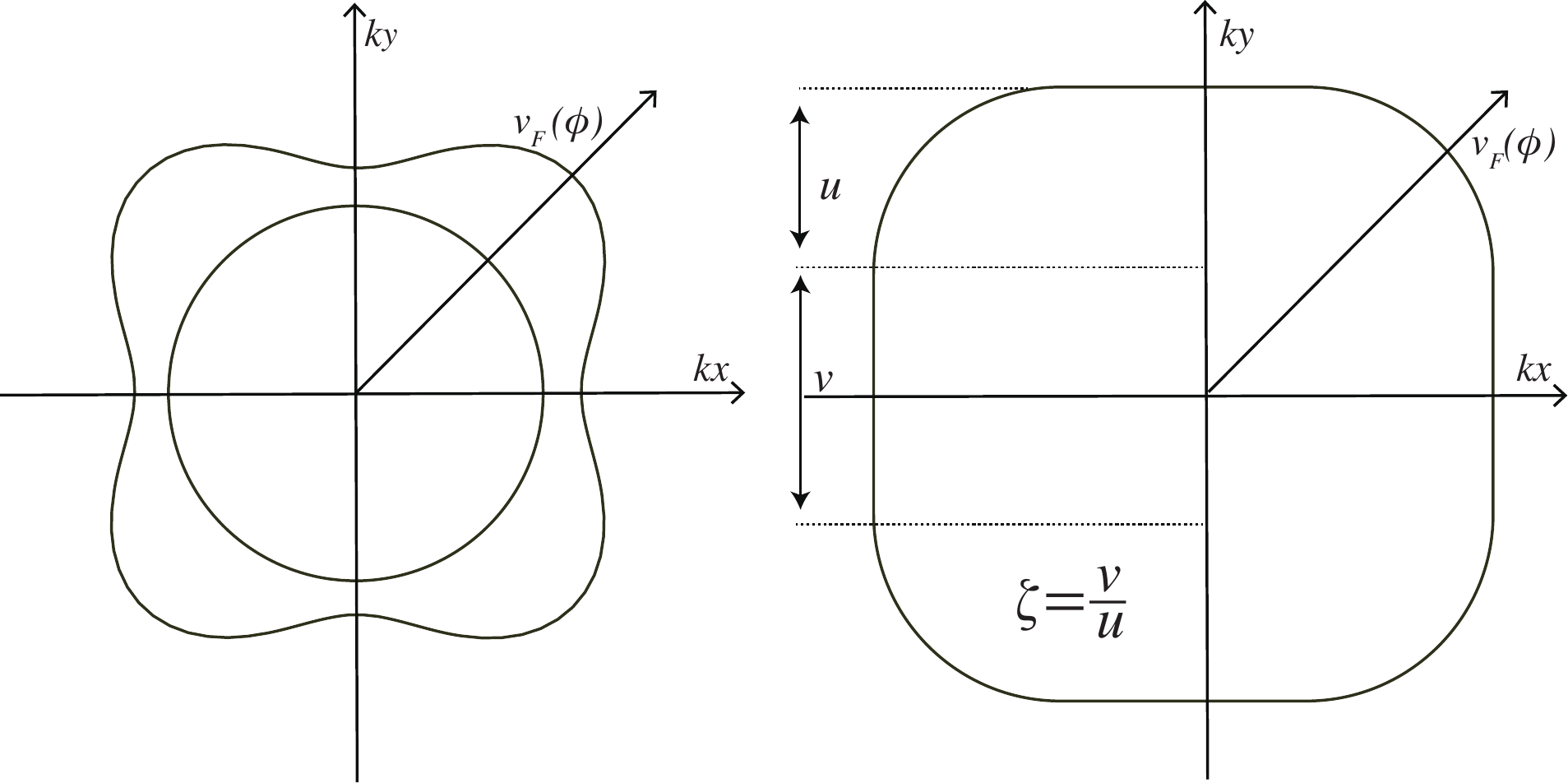}
\caption{
Schematic cross sectional views of the $b$-model (left) 
where on the Fermi circle the Fermi velocity $v_F(\phi)$ modulates sinusoidally.
The $\zeta$-model (right) where the Fermi surface is rectangular with the around portion $u$ and
the straight portion $v$ in the $ab$ plane.
}
\label{fig1-1}
\end{figure}

In order to capture the in-plane Fermi surface characteristics of three bands in Sr$_2$RuO$_4$,
we consider two types of in-plane Fermi surface models;
one is given by 
\begin{eqnarray}
v_F(\phi)=v_{F0}(1-b\cos 4\phi),
\label{vF}
\end{eqnarray} 
with the anisotropic parameter $b$ ($ > $0)~\cite{miranovic1}. The angle $\phi$ is measured from the $k_x$-axis or $a$-axis.
Let this be the $b$-model, a model to
design the $\gamma$-band whose Fermi surface is rather round and the Fermi velocity varies sinusoidally along the circle (see Fig.~\ref{fig1-1}).
We consider $b>0$ in the followings. It is a robust feature that the Fermi velocity $v^{\gamma}_F(\phi=0)$ is generically smaller than that
of $v^{\gamma}_F(\phi=\pi/4)$ because the Fermi surface of the $\gamma$-band is near the von Hove point at 
$(\pi, 0)$ in the Brillouin zone. For example, according to first principles band calculation\cite{suzuki},
$b=0.3\sim0.5$. Note, however, that the projected Fermi velocities on the (100) and (110) axes:
$\langle v_{(100)}^{\gamma}(\phi)^2 \rangle = \langle v_{(110)}^{\gamma}(\phi)^2 \rangle$
averaged over the Fermi velocity distribution Eq.~(\ref{vF}).

Since the Fermi surface shapes of the $\beta$-band and $\alpha$-band are square-like\cite{fermisurface}, 
we model it by the following $\zeta$-model\cite{udagawa}. As shown on the right panel of Fig.~\ref{fig1-1}, 
the Fermi surface consists of a 
parallel section with length $v$ and a round section $u$. We assume 
that the amplitude of the Femi velocity is constant everywhere. Thus the parameter $\zeta=v/u$
characterizes the squareness of the Fermi surface. $\zeta\rightarrow\infty$ ($\zeta$=0) corresponds to
a perfect square (a circle).

Since it is difficult to uniquely assign the parameters $b$ and $\zeta$ from band calculations,
they are presently only free parameters. However, we note that if the in-plane
gap function is isotropic, the in-plane $B_{c2}(\phi)$ anisotropy $\Gamma_{\phi}$ in the GL region is
given by 
\begin{eqnarray}
\Gamma_{\phi}\equiv{B_{c2}(\phi=0)\over B_{c2}(\phi={\pi\over 4})}=
\sqrt{\frac{\langle v_{(110)}(\phi)^2 \rangle}{\langle v_{(100)}(\phi)^2 \rangle}}=1
\label{unity}
\end{eqnarray} 
for the $b$-model, which is independent of the $b$ value, whereas
$\Gamma_{\phi}$ depends on the $\zeta$ value, for example, 
$\Gamma_{\phi}=1.06$ for $\zeta=1.0$, $\Gamma_{\phi}=1.13$ for $\zeta =2.0$, and $\Gamma_{\phi}=1.18$ for $\zeta=3.0$.
Thus we must be careful to choose the $\zeta$ value when considering various experimental situations. 
If the $\zeta$ value is too large, the constraint imposed by the experimental observation
of the absence of in-plane anisotropy is violated.
The observed in-plane anisotropy $\Gamma_{\phi}$ is very small near $T_c$ and is within 
at most 3$\%$ at $B=1$T\cite{kittakaBc2}.
We also note that $\Gamma_{\phi}>1$, which is contrary to the observation of $\Gamma_{\phi}<1$,
namely $B_{c2}(\phi=0)<B_{c2}(\phi={\pi\over 4})$ at lower temperatures~\cite{kittakaBc2}.
We will touch upon it in the last section.

\section{Horizontal line nodes}
\subsection{KPA results and Doppler shift picture}
We first introduce the four-fold oscillation amplitude $A_4(B)$, which is measured by field-rotating specific
heat experiments~\cite{kittaka0}, defined by
\begin{eqnarray}
A_4(B)\equiv{N(E=0, \phi={\pi\over4})-N(E=0, \phi=0)\over{N(E=0, \phi={\pi\over4})+N(E=0, \phi=0)}}
\label{A4def}
\end{eqnarray} 
with $N(E=0, \phi)\equiv N(\phi)$ being the zero energy DOS when the field is applied at angle $\phi$.

\begin{figure}[tbp]
\includegraphics[width=7cm]{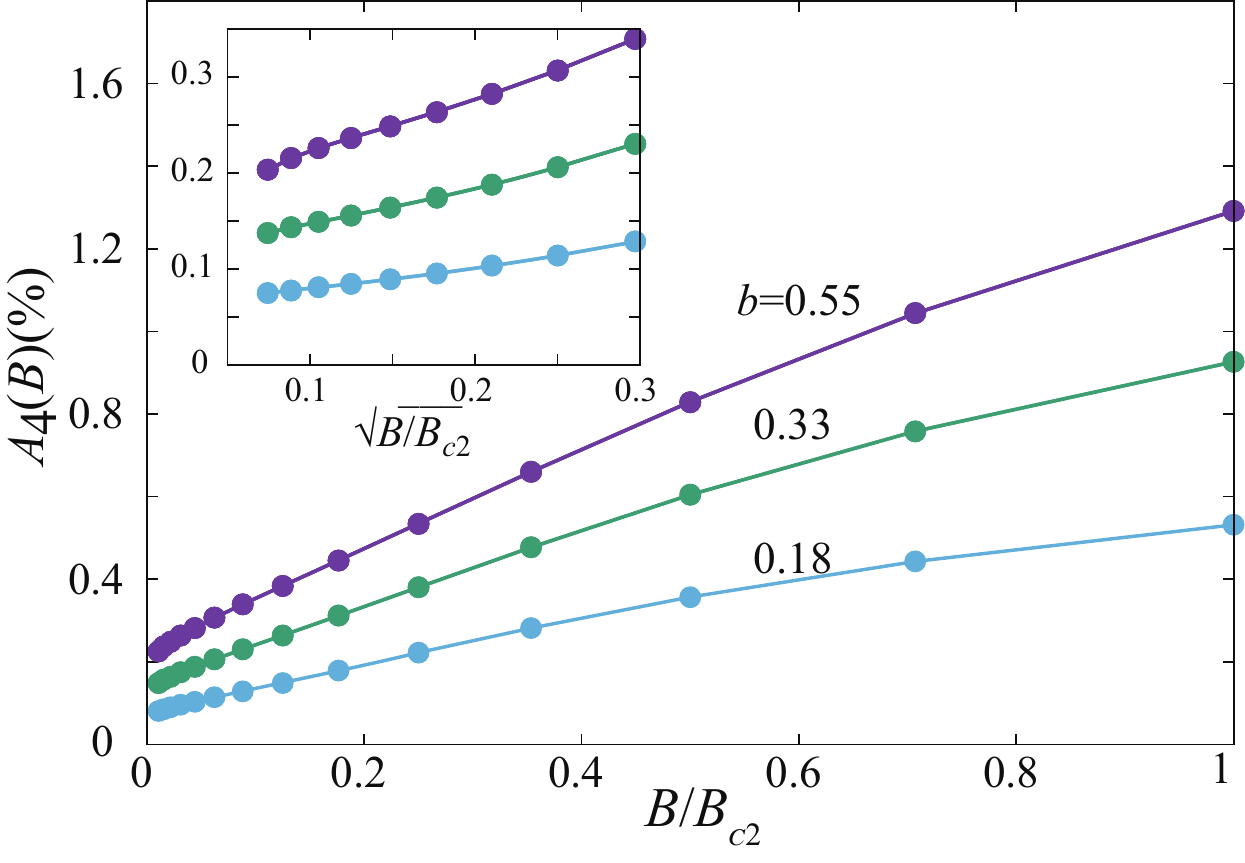}
\caption{(color online)
$A_4(B)$ for various $b$ values for the case of a 2D cylinder.
As $b$ increases $A_4(B)$ increases. The inset shows the low field parts of $A_4(B)$,
indicating the $A_4(B)\propto \sqrt B$ behavior.}
\label{fig2-2}
\end{figure}

We show the KPA results of $A_4(B)$ for HLN for two-dimensional (2D) cylindrical Fermi surface in Fig.~\ref{fig2-2}.
It is seen that $A_4(B)$ increases rather quickly 
which is approximately $A_4(B)\propto \sqrt B$ in lower fields as seen from the inset of Fig.~\ref{fig2-2}.
And it keeps increasing toward higher fields.
By increasing the Fermi velocity anisotropy $b$ introduced in Eq.~(\ref{vF}) the amplitude $A_4$ grows.
The growing rate is linear in $b$ at least for smaller and moderate $b$ values.

Namely, we see

 (1) $A_4(B) >0$, 
 
 (2) $A_4(B)$ monotonically increases, and
 
(3) $A_4(B)$ approaches a finite value as $B\rightarrow0$.

Some of the findings are understood in terms of the semiclassical Doppler shift picture as follows:
In the presence of linear line nodes in general, the average total density of states $N(E)$ 
has a V-shaped energy dependence 
for all $B$ values from $B=0$ up to $B_{c2}$\cite{nakai}.
The energy $E_D$ associated with the Doppler shift is given by
$E_D \propto {\bf v}_s\cdot {\bf v}_F({\bf p})$ for the quasi-particles
propagating along the ${\bf p}$-direction\cite{volovik,hirschfeld}. Thus $E_D$ depends on the 
field direction through $v_F(\phi)$. 
Under field rotation $E_D(\phi)$ oscillates proportional to $v_F(\phi)$.




\begin{figure}[tbp]
\includegraphics[width=8cm]{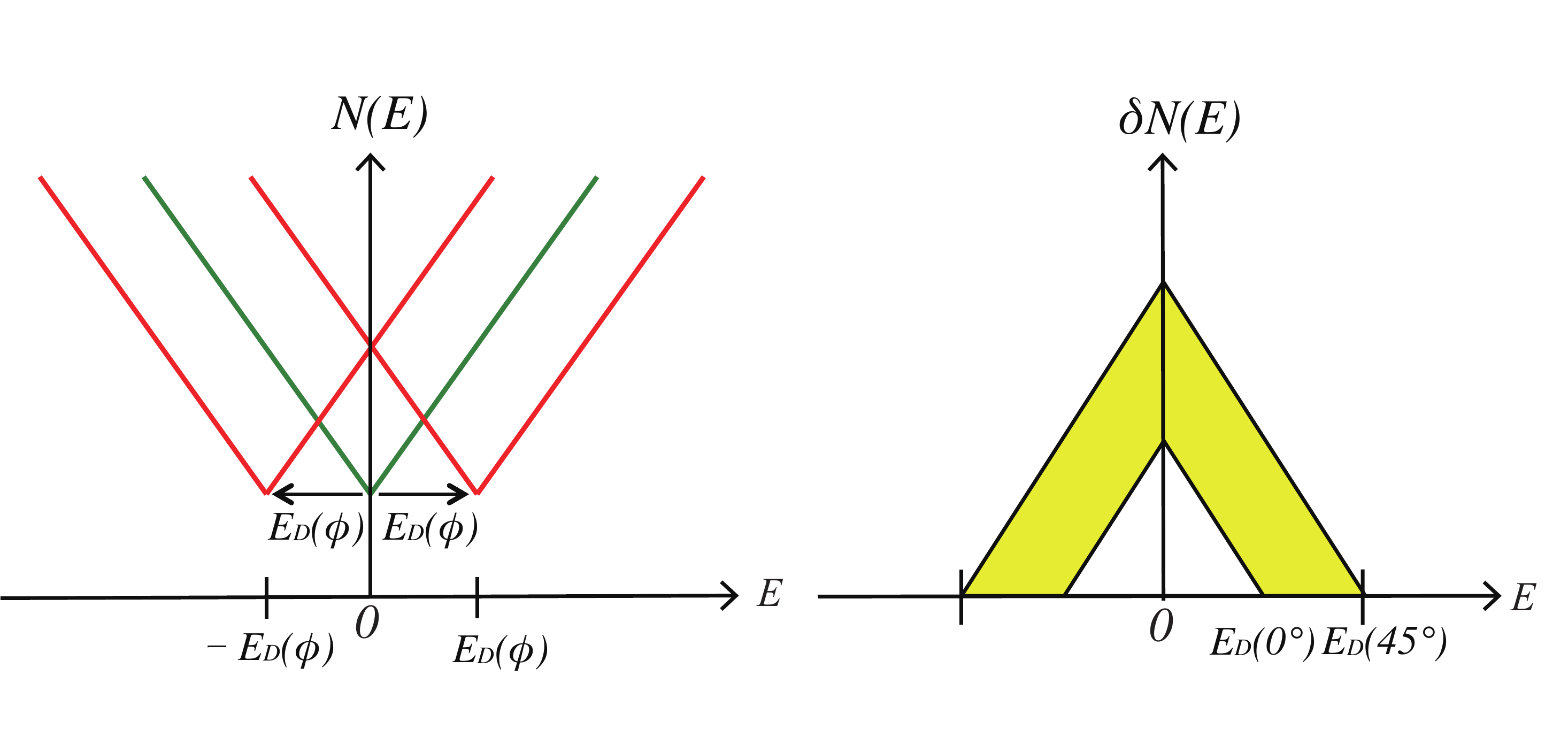}
\caption{(color online)
Schematic pictures of the Doppler shift.
Left: Original V-shape DOS (green) is shifted to two DOS (red) by $\pm E_D(\phi)$, producing the increment in ZDOS.
Right: Increment $\delta N(E)$ of DOS due to the Doppler shift as a function of $E$ obtained by subtracting the two
shifted DOS (red) from original DOS (green). A triangle centered at $E=0$ forms whose size depends 
on the field-orientation $\phi$ through $E_D(\phi)$.
}
\label{fig2-6}
\end{figure}

As schematically illustrated in Fig.~\ref{fig2-6}, the increment $\delta N(E)$ of the DOS by the Doppler shift appears 
only at around $E=0$ as a triangular area centered at $E=0$. The area of this triangle is 
proportional to $v_F(\phi)$ which gives rise to the DOS oscillation. Thus 
\begin{eqnarray}
A_4(B)\propto E_D(B)\cdot{dN(E=+0)\over dE}
\label{A4}
\end{eqnarray} 
with ${dN(E=+0)\over dE}=N'(E=+0)$ is the slope of $N(E)$ near $E=+0$.
Since $E_D(B)$ is an increasing function of $B$ through the ${\bf v}_s$ factor,
$A_4(B)$ increases with $B$ and is proportional to $b$.
This idea based on the Doppler shift effect is consistent with some aspects of the 
KPA results. 




For the 2D cylinder FS
$N(E)$ changes a V-shape at lower energy to a U-like shape 
as $|E|$ increases (see $B=0$ curve in Fig.~\ref{fig2-9ab}(a)), thus $A_4(B)$ keeps increasing as the field
strength is increased.

Although such $A_4(B)$ behavior in the KPA supports the Doppler shift picture,
it should be noticed that this simple $N'(E=+0)$ behavior 
must be more carefully reexamined as will be seen shortly.
The shortcoming of the KPA based on the single vortex approximation is apparent 
because the effects of vortex core overlapping become crucial at mid and higher fields.

\begin{figure}[tbp]
\includegraphics[width=7cm]{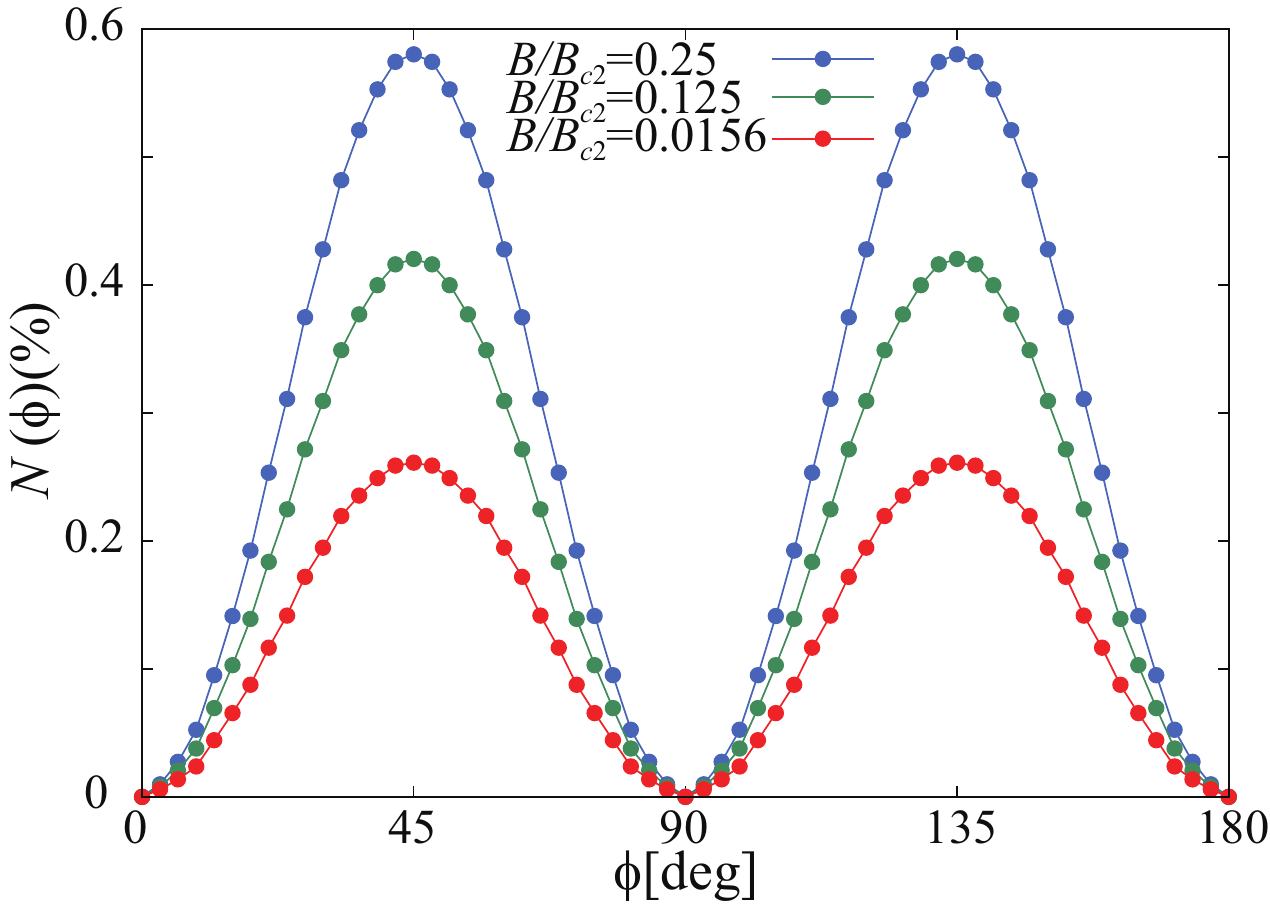}
\caption{(color online)
Oscillation patterns of $N(\phi)$ in the $b$-model for several $B$. $b=0.33$.}
\label{fig2-7-1}
\end{figure}

\begin{figure}[tbp]
\includegraphics[width=7cm]{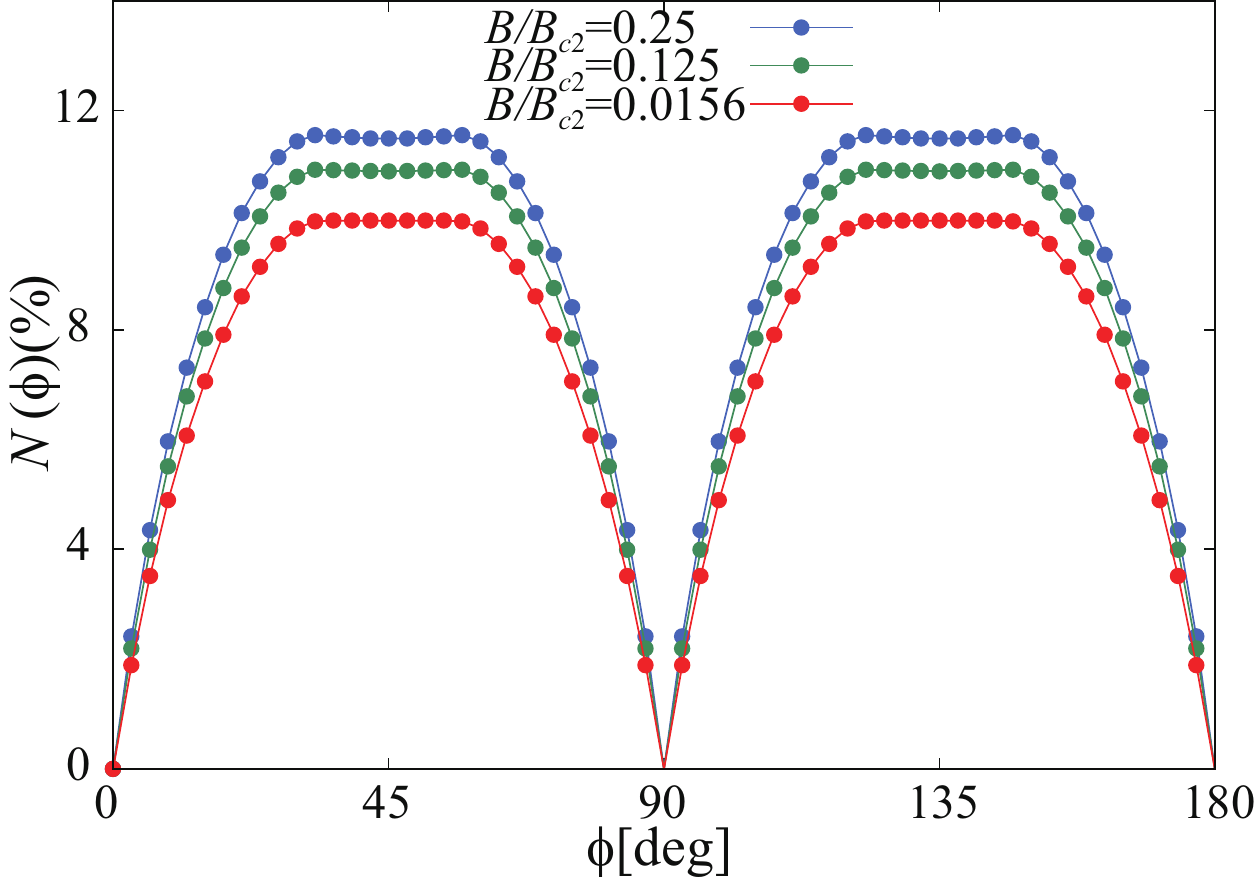}
\caption{(color online)
Oscillation patterns of $N(\phi)$ in the $\zeta$-model for several $B$ values. $\zeta=1.0$.}
\label{fig2-7-2}
\end{figure}

\begin{figure}[tbp]
\includegraphics[width=7cm]{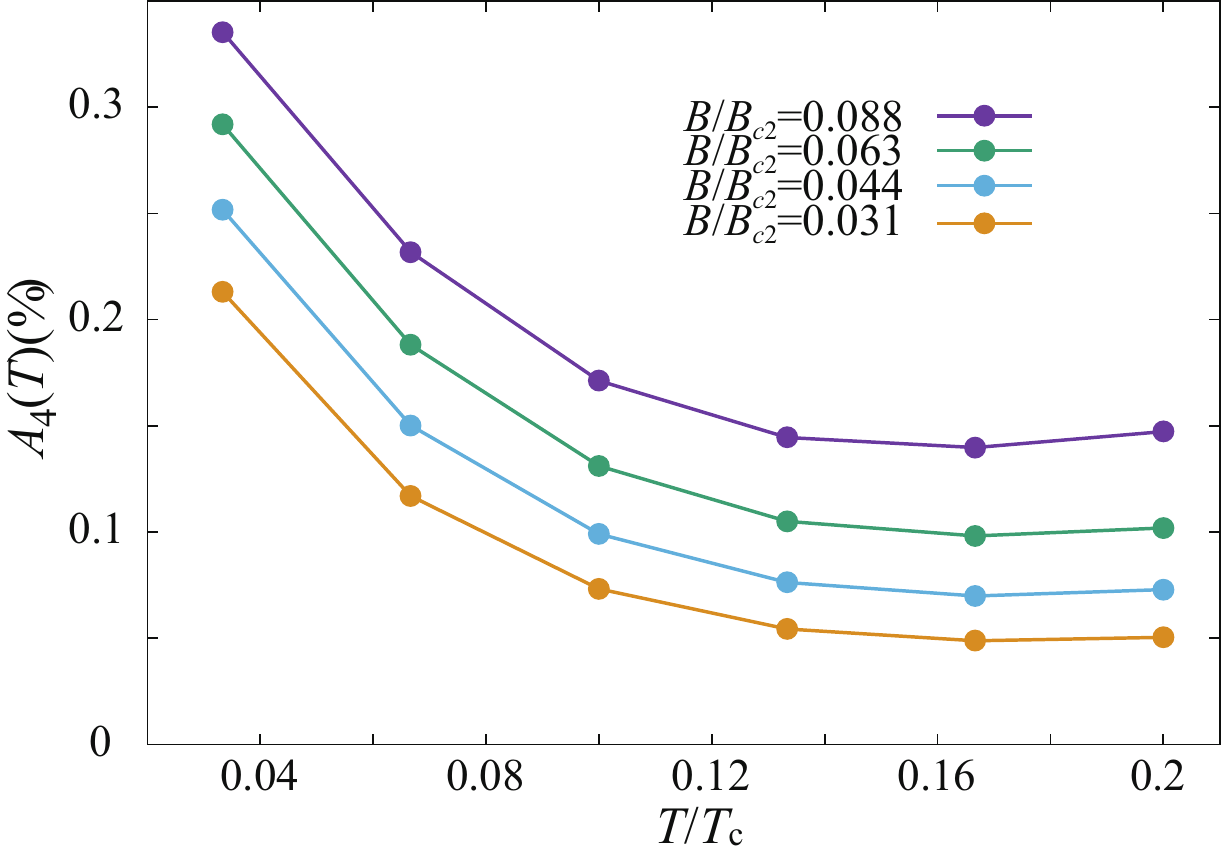}
\caption{(color online)
Temperature dependences of $A_4(T)$ for various $B$ values,
showing that $A_4(T)$ quickly diminishes as $T$ grows. $b=0.33$
for the $b$-model.}
\label{fig2-4}
\end{figure}

As seen from Figs. \ref{fig2-7-1} and \ref{fig2-7-2}, which show the results of the oscillation patterns of 
$N(\phi)$ for the $b$-model and $\zeta$-model, the same general oscillation trend is seen, i.e.,
the (100) minimum, or $\phi=0$. The oscillation patterns sensitively reflect the FS shape.
As $\zeta$ increases or the FS shape becomes rectangular, the oscillation patterns are distorted far from a simple sinusoidal form 
as seen in the $b$-model cases.

The $T$ dependence of $A_4(T)$ defined by
\begin{eqnarray}
A_4(T)\equiv{C(T, \phi=\pi/4)-C(T, \phi=0)\over{C(T, \phi=\pi/4)+C(T, \phi=0)}}
\label{A4T}
\end{eqnarray} 
is evaluated through the specific heat formula:
\begin{eqnarray}
{C(T)\over T}=\int_0^{\infty} {dE \over T}{E^2\over 2T^2}{N(E) \over \cosh^2({E\over 2T})}.
\label{C(T)}
\end{eqnarray} 
As shown in Fig.~\ref{fig2-4}, the $T$ dependence of $A_4(T)$ is also consistent with the 
Doppler shift picture because the increment $\delta N(E)$ of DOS 
by the Doppler shift is confined to being near the $E=0$ energy region,
as indicated by the triangle in Fig.~\ref{fig2-6},
meaning that $A_4(T)$ is also limited to a low $T$ region.

In summary of this subsection,
we explained the physics of the Doppler shift picture for describing the DOS oscillation.
It is likely that $A_4(B,T)$ of the in-plane DOS oscillation is positive for HLN, namely
\begin{eqnarray}
A_4(B,T)\geqq0.
\end{eqnarray} 

\subsection{Full Eilenberger calculations without PPE}
Having established the applicability of the Doppler shift picture through the results derived by KPA for 
the Eilenberger equation, we proceed
further by more accurately solving the full Eilenberger equation self-consistently 
under the realistic situation, namely the cylindrical Fermi surface model for
Sr$_2$RuO$_4$ with and without Pauli paramagnetic effect.
The gap structure with the horizontal line nodes is written as
\begin{eqnarray}
\Delta(k)=\Delta_0\cos ck_z
\end{eqnarray} 
with $c$ being the lattice constant along the $c$-axis.
The other parameters are the same as before \cite{amano1,amano2} except for the in-plane anisotropic Fermi velocity,
which is modeled by the $b$-model.

\begin{figure}[tbp]
\includegraphics[width=7cm]{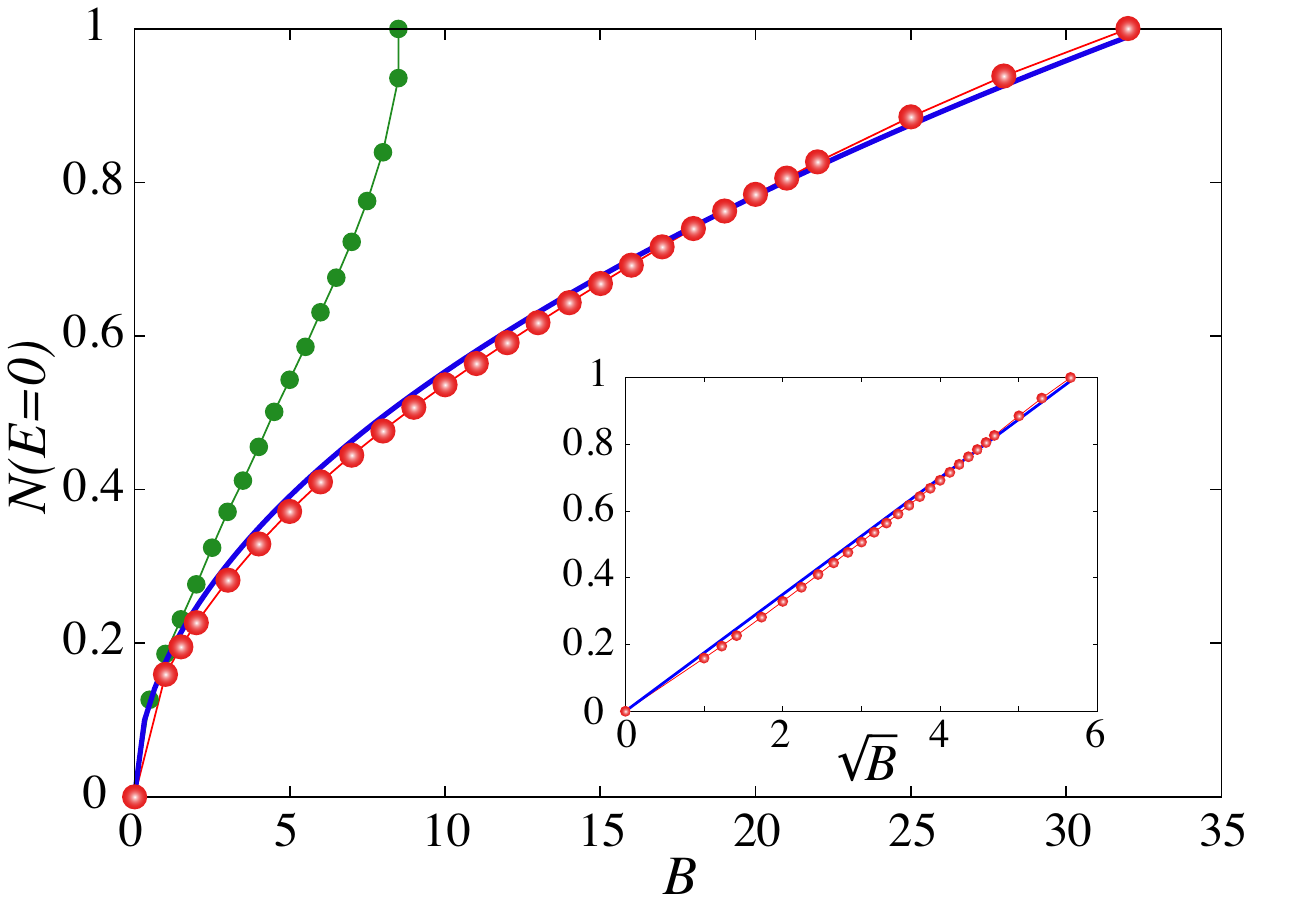}
\caption{(color online)
ZDOS $N(E=0)$ as a function of $B$  for $\mu=0$ (red dots)
and $\mu=0.04$ (green dots). The blue line along the red dots 
indicates $N(E=0)=\sqrt {B/B_{c2}}$
with $B_{c2}=32$. The inset shows the $\sqrt B$ plot.}
\label{fig2-14}
\end{figure}

The calculated field dependent zero energy DOS $N(E=0)$ normalized by the normal state value $N_0$ is shown as red dots
in Fig.~\ref{fig2-14}.
It is seen that $N(E=0)$ is a typical form $\sqrt B$ characteristic to the nodal gap structure\cite{IchiokaPara,Machida214}.
In fact as compared with $N(E=0)=\sqrt {B/B_{c2}}$ curve, the numerical points are described remarkably
well by this formula, not only at lower $B$ which is expected to be valid, but also all the way up to $B_{c2}$.  

\begin{figure}[tbp]
\includegraphics[width=7cm]{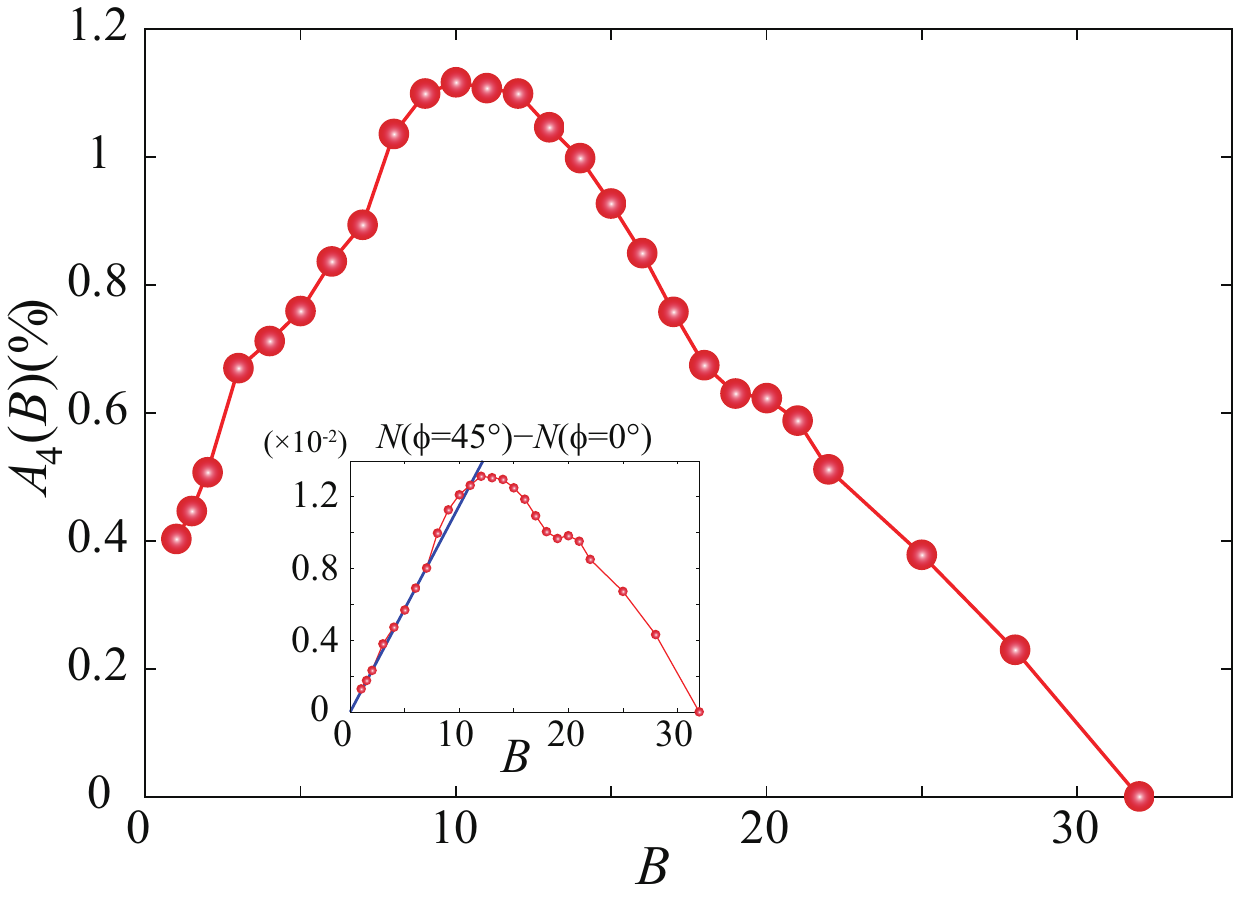}
\caption{(color online)
$A_4(B)$ from the fully self-consistent solution of Eilenberger theory for HLN. $b=0.2$ and $\mu=0$ (red dots).
The inset shows the ZDOS difference $N(\phi=45^{\circ})-N(\phi=0^{\circ})$
as a function of $B$. At lower fields it is linear in $B$.}
\label{fig2-8}
\end{figure}

The angle dependent oscillation amplitude $A_4(B)$ is also calculated in Fig.~\ref{fig2-8} shown as red dots.
This result shows:

 (1) As $B\rightarrow 0$ $A_4(B)$ tends to a finite value.

(2) $A_4(B)$ exhibits a maximum around $B_\textrm{max}/B_{c2}\cong 10/32 \sim0.3$.

(3) After it maximizes, $A_4(B)$ decreases almost linearly as $B\rightarrow B_{c2}$(=32).

(4) $A_4(B)\geq 0$, i.e., $A_4(B)$ is positive. 

\begin{figure}[tbp]
\includegraphics[width=8cm]{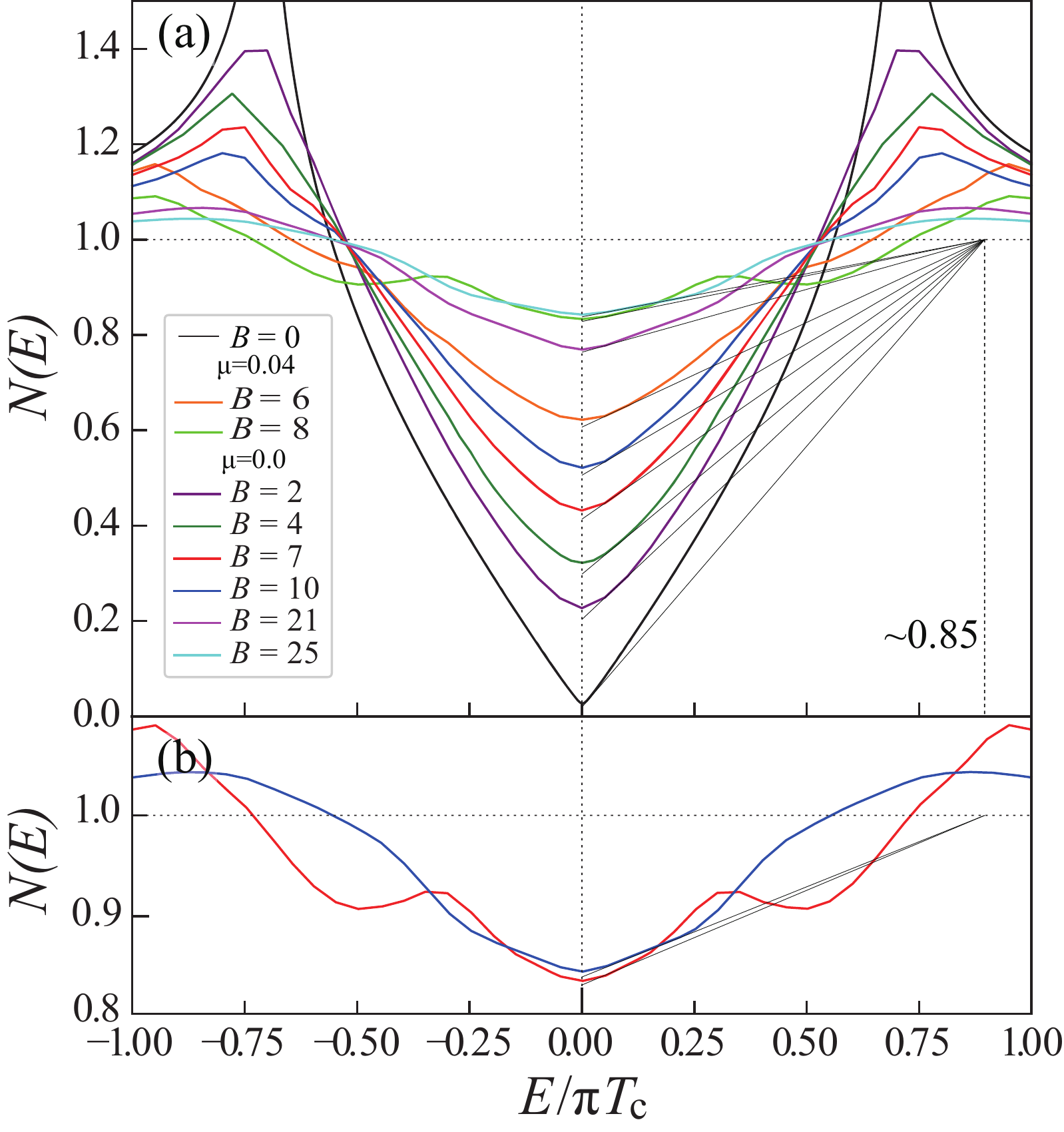}
\caption{(color online)
(a) $N(E)$ obtained by solving the full Eilenberger theory self-consistently for various fields.
Both $\mu=0$ and $\mu=0.04$, including the 2D DOS $N(E)$ for $B=0$.
The slopes at $E\sim +0$ have a common focal point at $N(E/\pi T_{\rm c}=0.85)=1$,
demonstrating the DOS scaling: $N'(E\sim +0)\propto 1-N(E=0)$.
(b) Detailed comparison of the slopes for top two curves in (a) with $\mu=0$ (blue) and $\mu=0.04$ (red).
They have almost same slopes near $E=+0$,
but at higher energies are widely different due to PPE.
Note that due to numerics $N(E)$ deviates slightly at $E=0$ from ideal V-shape form.
}
\label{fig2-9ab}
\end{figure}

Result (1) coincides with that from KPA mentioned above.
However, (2), (3), and (4) are not covered by KPA, simply because of inherent limitations
due to the single vortex approximation in KPA.
Thus, the Full Eilenberger calculation adds the new features (2), (3), and (4). 
In order to understand the physical origin of the new features and 
further refine the Doppler shift picture, we have carried out  extensive computations.
We uncover several novel facts that were crucial in
determining the $A_4(B)$ behavior.
As shown in Fig.~\ref{fig2-9ab}, the total DOS $N(E)$ averaged over the spatial points 
within the vortex unit cell forms a characteristic V-shape near $E=0$,
including when $B=0$. Because the value of $N(E)$ at $E=0$ is sensitive to numerical error,
the obtained DOS $N(E)$ is somewhat approximated and rounded near $E\sim 0$, but retains an approximate
V-shape (see Ref.~\onlinecite{nakai} for details).
The opening angle of the V-shape depends on $B$, namely it becomes shallower as $B$ 
increases.

\begin{figure}[tbp]
\includegraphics[width=7cm]{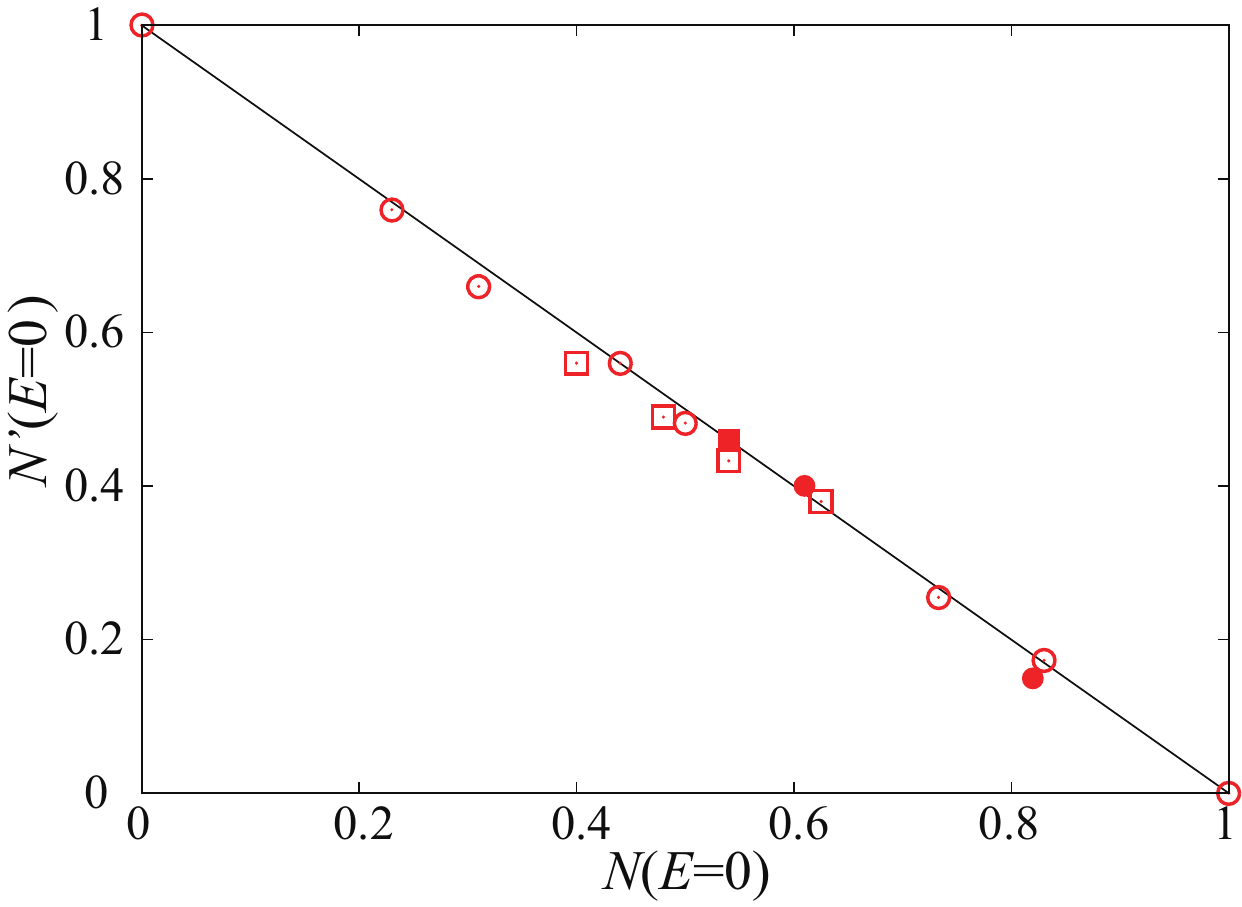}
\caption{(color online)
Slopes  $N'(E\sim +0)$ normalized by the N(E=0) case at $B=0$ are plotted as a function of $N(E=0)$.
The data is extracted from Fig.~\ref{fig2-9ab} and also includes data 
from Nakai \textit{et al.}~[\onlinecite{nakai}] (empty squares) and Ichioka \textit{et al.}~[\onlinecite{IchiokaPara}] (filled circles).}
\label{fig2-10}
\end{figure}

As $B\rightarrow B_{c2}$ the derivative $[{dN(E)\over dE}]_{E\simeq +0}=N'(E\sim+0)$
 at lower energy continuously decreases and tends to vanish at $B=B_{c2}$.
 We find a focal point of the tangential lines,
meaning that $N'(E\sim+0, B)$ is a linear function $N(E=0,B)$.
As shown in Fig.~\ref{fig2-10}, we find the DOS scaling law:
\begin{eqnarray}
{N'(E\sim+0, B)\over N'(E\sim+0, B=0)}=1-N(E=0,B).
\label{dosscaling}
\end{eqnarray} 

This simple relationship includes the Pauli limiting cases with $\mu\neq0$
(the filled points in Fig.~\ref{fig2-10}).
By substituting Eq.(\ref{dosscaling}) into Eq. (\ref{A4}), we obtain
\begin{eqnarray}
A_4(B)\propto E_D(B)(1-N(E=0,B)).
\end{eqnarray} 

To determine the field dependence of the Doppler shift energy $E_D(B)$,
we evaluate the numerator of Eq. (\ref{A4def}), namely the difference in the ZDOS $N(\phi=45^{\circ})-N(\phi=0^{\circ})$
from the Eilenberger full solutions.
As seen from the inset of Fig.~\ref{fig2-8}, the difference in the ZDOS in $B$ is linear 
at lower fields. On the other hand, the denominator of Eq. (\ref{A4def}): $N(\phi=0^{\circ})+N(\phi=45^{\circ})\propto \sqrt{B}$
due to the Volovik effect~\cite{volovik}.
The resulting $A_4(B)\propto\sqrt{B}$ in Eq.(\ref{A4def}) at lower fields.
The linearity in the difference of the ZDOS can be understood as follows:
the extended quasi-particle contributions proportional to $\sqrt{B}$ cancel out, but the core localized 
quasi-particle contributions remain and give rise to the oscillation whose field dependence is
obviously proportional to the flux number or $B$. Thus this is a contributing factor to the DOS oscillation
at lower fields that vanishes at higher fields when the core localized quasi-particles overlap
each other.

We can estimate this field $B_\textrm{max}$ by calculating the field
at which the elongated vortex cores start overlapping.
$B_\textrm{max}/B_{c2}^{ab}=\xi_c/\lambda=1/\kappa=1/2.7$
with the GL parameter $\kappa$ along the $c$-axis chosen to be 2.7
in our calculations~\cite{amano1,amano2} as mentioned before. 
This agrees well with the numerical calculation shown in Fig.~\ref{fig2-8}.

In view of the above DOS scaling we postulate that $A_4(B)$ is determined uniquely by $N(0)$
and extend the DOS scaling, including $A_4(B)$.
  
\begin{figure}[tbp]
\includegraphics[width=8cm]{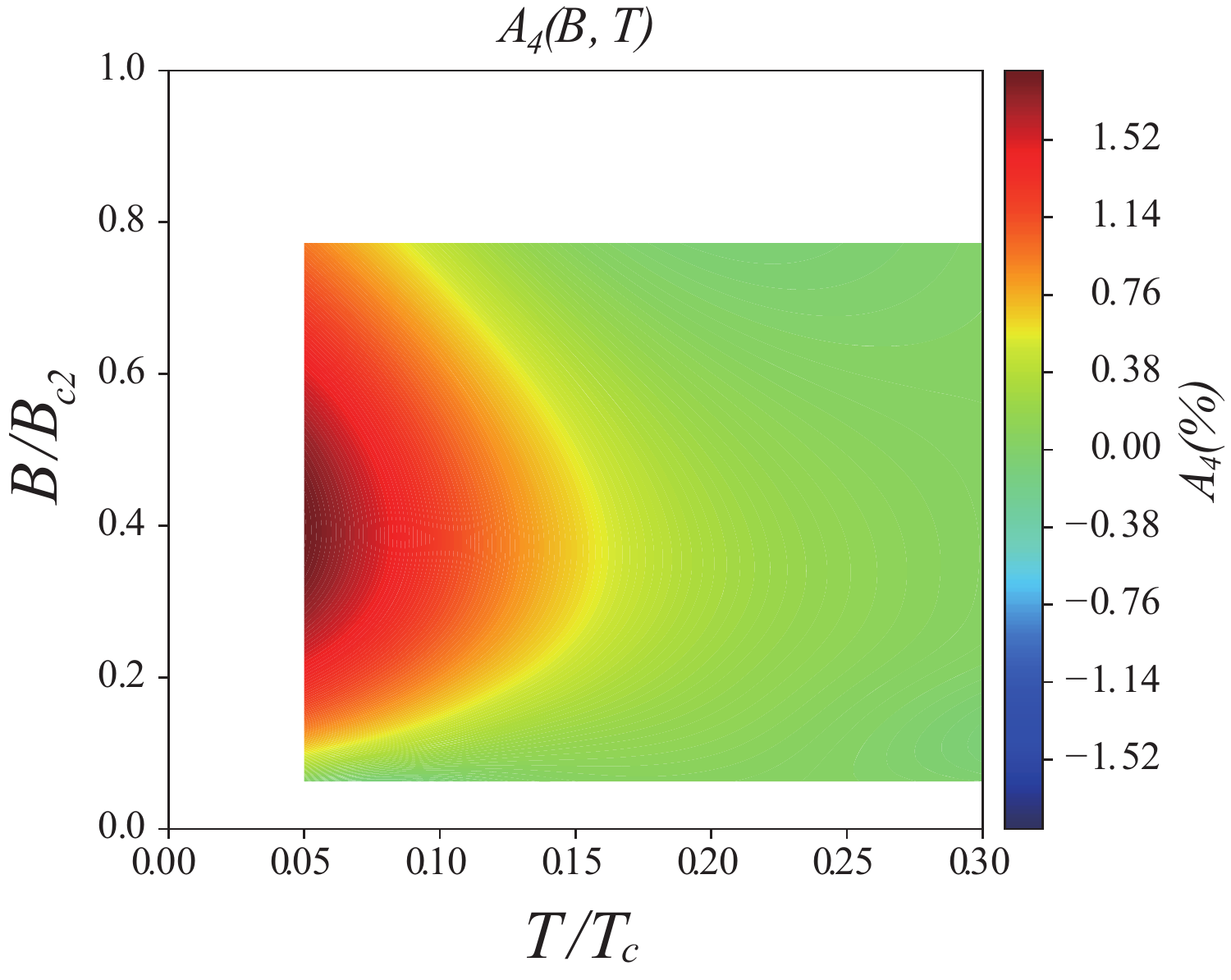}
\caption{(color online)
Landscape of $A_4(B,T)$ for $b=0.5$ and $\mu=0$.}
\label{fig2-12}
\end{figure}

Once the DOS $N(E)$ is calculated, it is easy to evaluate the specific heat $C(T)$ by Eq. (\ref{C(T)}),
after which  $A_4(T)$ can be calculated using Eq. (\ref{A4T}).
The obtained  $A_4(B,T)$ is illustrated in Fig.~\ref{fig2-12} as a contour map.
It is seen that the landscape is simple:
A hill in the $B$-$T$ plane is situated at low $T$ and $B_\textrm{max}/B_{c2}\sim0.3$
where a ridge extends toward higher temperatures.
We notice that this hill structure is confined to the low temperature region
only up to at most $\sim0.25T_c$. This 
is contrasted with the vertical line node case where the 
$B$-$T$ landscape is much more complicated, exhibiting  an $A_4$ sign change region,
local maximum and minimum, and $A_4$ is a finite up to at least 
$\sim 0.4T_c$ as seen from Fig~\ref{fig2-13-2}(b) (also see Figs. 11, 12 and 13  in Ref.[\onlinecite{hiragi}]).

\subsection{Full Eilenberger calculation with PPE and DOS scaling}
We performed the same Eilenberger computations 
by taking into account the Pauli paramagnetic effects (PPE) with $\mu=0.04$.
The ZDOS $N(E=0)$ as a function of $B$ is shown in Fig.~\ref{fig2-14} as green dots.
Due to the strong PPE the system exhibits a first order transition at $B_{c2}=8.5$,
indicated by a jump of $N(E=0)$.
The presence of HLN is recognized a prominent $\sqrt B$ behavior at lower fields,
but it is modified strongly in the middle and high field regions due to PPE.
This behavior is consistent with previous calculations\cite{Machida214}.

\begin{figure}[tbp]
\includegraphics[width=7cm]{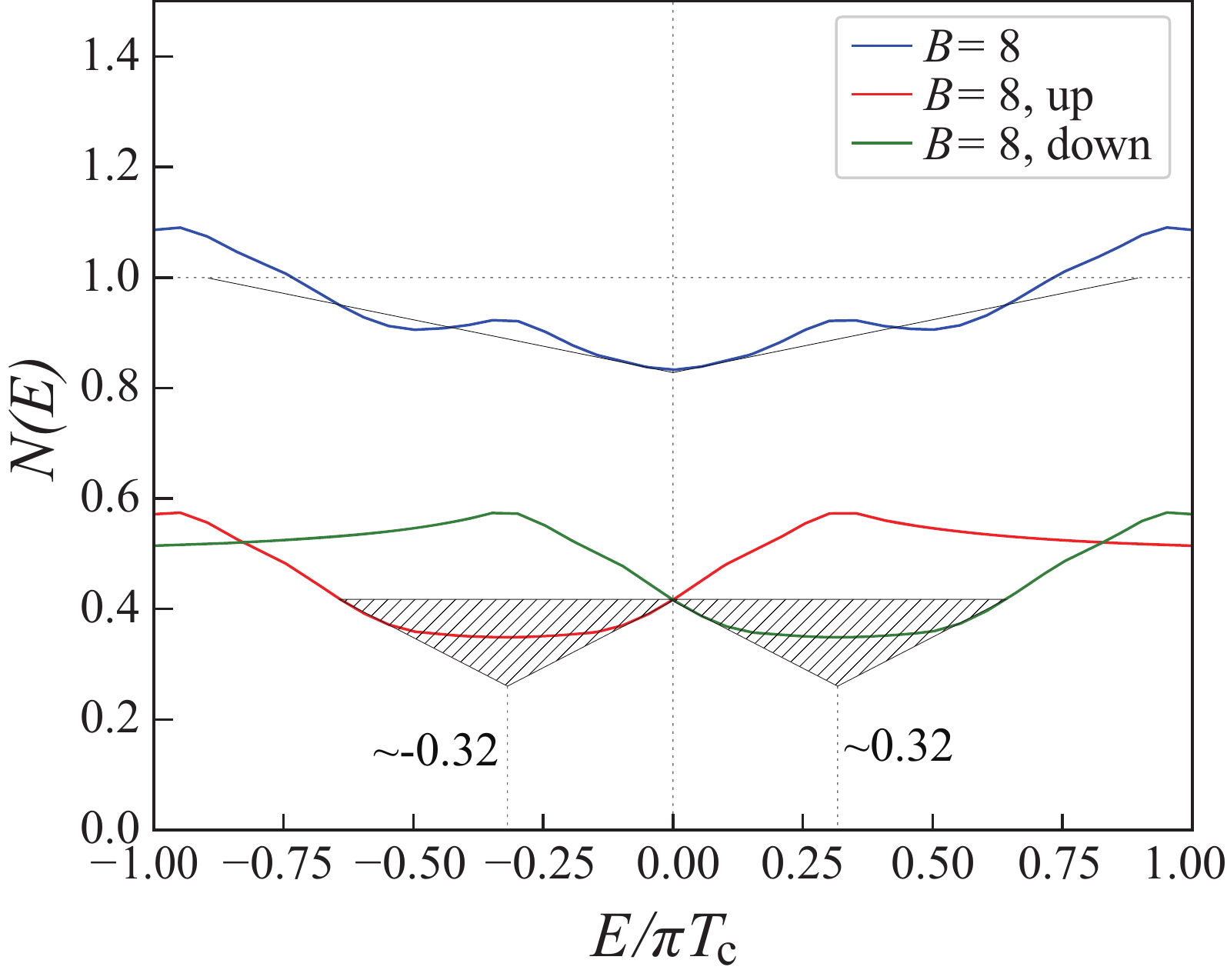}
\caption{(color online)
Reconstruction of the V-shape DOS $N(E)$ under PPE.
Red and green curves are the spin-resolved DOS and the blue curve is total DOS.
To accommodate the excess Pauli paramagnetism due to PPE, the original
V-shape $N(E)$ for spin-up and spin-down are modified to have flat bottoms
shown by shaded triangles. Note that in spite of this modification the slope of 
the original V-shape DOS is preserved under PPE. See the details in the main text.
}
\label{fig2-15}
\end{figure}

\begin{figure}[tbp]
\includegraphics[width=6cm]{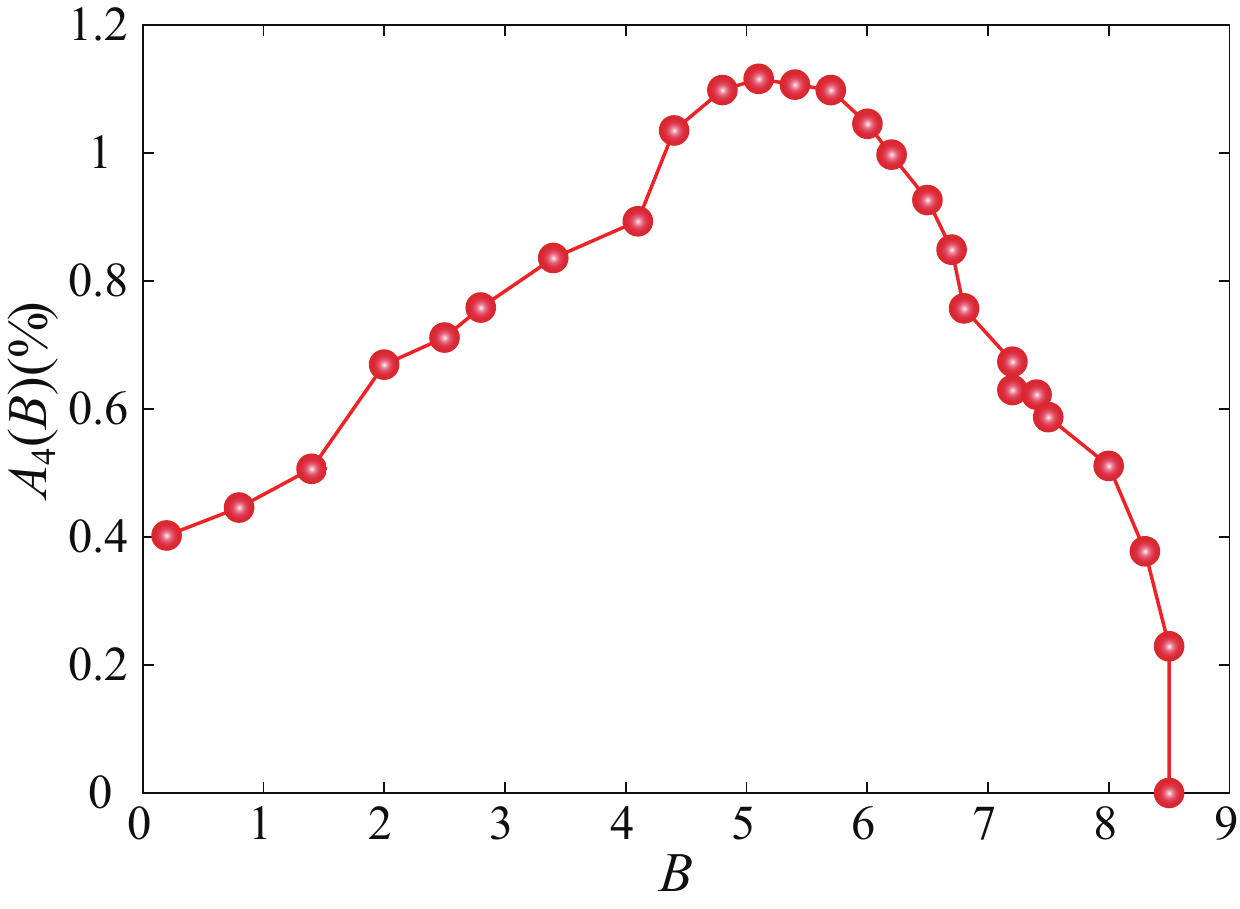}
\caption{(color online)
$A_4(B)$ under PPE obtained by scaling $A_4(B)$ in Fig.~\ref{fig2-8}.
}
\label{fig2-11}
\end{figure}

The DOS $N(E)$ is also calculated to estimate $A_4(B,T)$ under the PPE influence
by using the DOS scaling.
As shown in Fig.~\ref{fig2-15}, the total DOS $N(E)$ is decomposed into the spin-up and spin-down 
components, which are Zeeman split due to PPE.
In order to accommodate the induced Pauli paramagnetic component,
the original V-shaped DOS is reshaped as seen from the guided V-shape lines in Fig.~\ref{fig2-15}.
Namely the bottoms of the Zeeman shifted DOS curves become flat as seen from the red and green curves in in Fig.~\ref{fig2-15}
or ideally completely flat indicated by the shaded inverted triangles.
Their areas are exactly equal to the particle number corresponding to the induced paramagnetic moment.
Because of this flatness the resulting total DOS curve keeps the original V-shape
with the same slope $N'(E\sim+0)$ as that without PPE.
This slope and others\cite{nakai,IchiokaPara} are 
plotted in Fig.~\ref{fig2-10} as the filled symbols which are all embedded in  the points without PPE.
Note that as shown in Fig.~\ref{fig2-9ab}(b) the two V shaped DOS's with and without PPE
have the almost same slopes when their $N(E=0)$ are same.
Therefore, we establish a general DOS scaling Eq.~(\ref{dosscaling}):
$N'(E\sim+0)\propto 1-N(E=0)$ again. This time we include the PPE.

It is not difficult to estimate $A_4(B,T)$ under PPE by applying DOS scaling:
starting with $A_4(B)$ without PPE in Fig.~\ref{fig2-8},
then $A_4(B)$ under PPE is obtained by using the correspondence that 
the same $N(E=0)$ yields the same $A_4(B)$, which is displayed in Fig.~\ref{fig2-11}.

\begin{figure}[tbp]
\includegraphics[width=7cm]{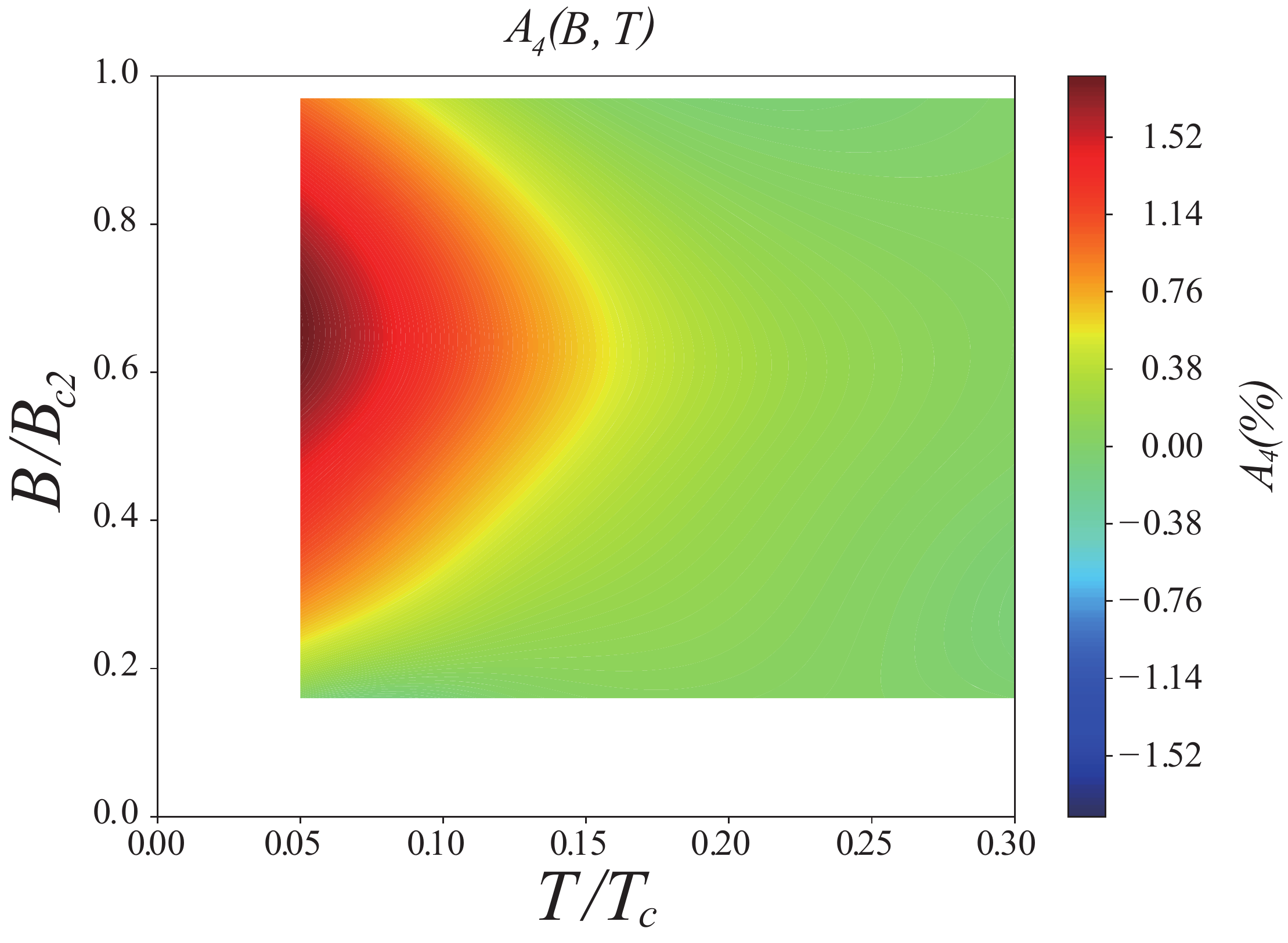}
\caption{(color online)
Landscape of $A_4(B,T)$ for $b=0.5$ and $\mu=0.04$ obtained by using the DOS scaling 
from Fig.~\ref{fig2-12}.
}
\label{fig2-13}
\end{figure}

We had applied the same DOS scaling to construct the $A_4(B,T)$ contour map for Fig.~\ref{fig2-12}.
The $A_4(T)$ data for a given $B$ without PPE was transformed to that under PPE.
The  same procedure was carried out for the $A_4(B)$ data from in Fig.~\ref{fig2-11}.
The result  is depicted in Fig.~\ref{fig2-13} where again the landscape is simple without any sign change
region. The ridge is now situated at around $B_\textrm{max}/B_{c2}\sim0.6$ -- $0.7$;
this slightly higher $B$ in comparison to that in Fig.~\ref{fig2-12} is due to PPE.
 
 Here we notice that comparable full Eilenberger calculations with and without PPE for vertical
 line node case with same cylindrical Fermi surface model ($\Gamma =60$) are
 done before\cite{hiragi}. The obtained $A_4(B,T)$ landscapes are quite different from that of the HLN
 cases and will be shown later in Fig.~\ref{fig2-13-2}(b).
  
\subsection{Multiband consideration}
We have discussed the angle-resolved DOS in terms of the $b$-model, which models the $\gamma$ band.
As for the $\beta$ band we apply the $\zeta$-model.
As seen from Figs.  \ref{fig2-7-1} and \ref{fig2-7-2}, 
the oscillations are qualitatively similar, i.e., both exhibit a (100) minimum,
though the oscillation patterns are different.
Because both oscillations are the same sense, the total $A_4(B, T)$, which is given by adding up two
contributions as a zeroth approximation.
Thus the conclusion that $A_4(B, T)$ is positive for the $B$ and $T$ plane
remains unchanged even under the multiband effect.
It is reasonable to expect that $A_4(B)$ in Figs.~\ref{fig2-8} and \ref{fig2-11} do not change in the essential way even 
taking into account the multiband effect into the microscopic Eilenberger calculation. 
Those will be contrasted with the vertical line node cases as seen shortly.

\section{Vertical line nodes}
It is known that when the vertical line nodes (VLN) are present on the $\gamma$ band,
which is well described by the $b$-model, $A_4(B,T)$ exhibits the sign change
both as functions of $B$ and $T$ at around $B_{ch}/B_{c2}\sim0.35$ and $T_{ch}/T_{c}\sim0.15$~\cite{hiragi}.
The (100) minimum of $A_4(B,T)$ is realized in low (high) field and at low (high) temperatures
for the $d_{x^2-y^2}$ ($d_{xy}$) symmetry case.
Therefore, it is obvious that this is not the case for Sr$_2$RuO$_4$.
Here we focus on the $\zeta$-model corresponding to the $\beta$ band whose 
$A_4(B,T)$ behavior is not yet fully analyzed. 
We calculate $A_4(B)$ and $A_4(T)$ for the VLN cases with KPA. 
We confirm that  results in KPA are basically consistent with the full Eilenberger 
calculations done before\cite{hiragi} for $\zeta$=0.

\subsection{$d_{x^2-y^2}$-symmetry}
\begin{figure}[tbp]
\includegraphics[width=6cm]{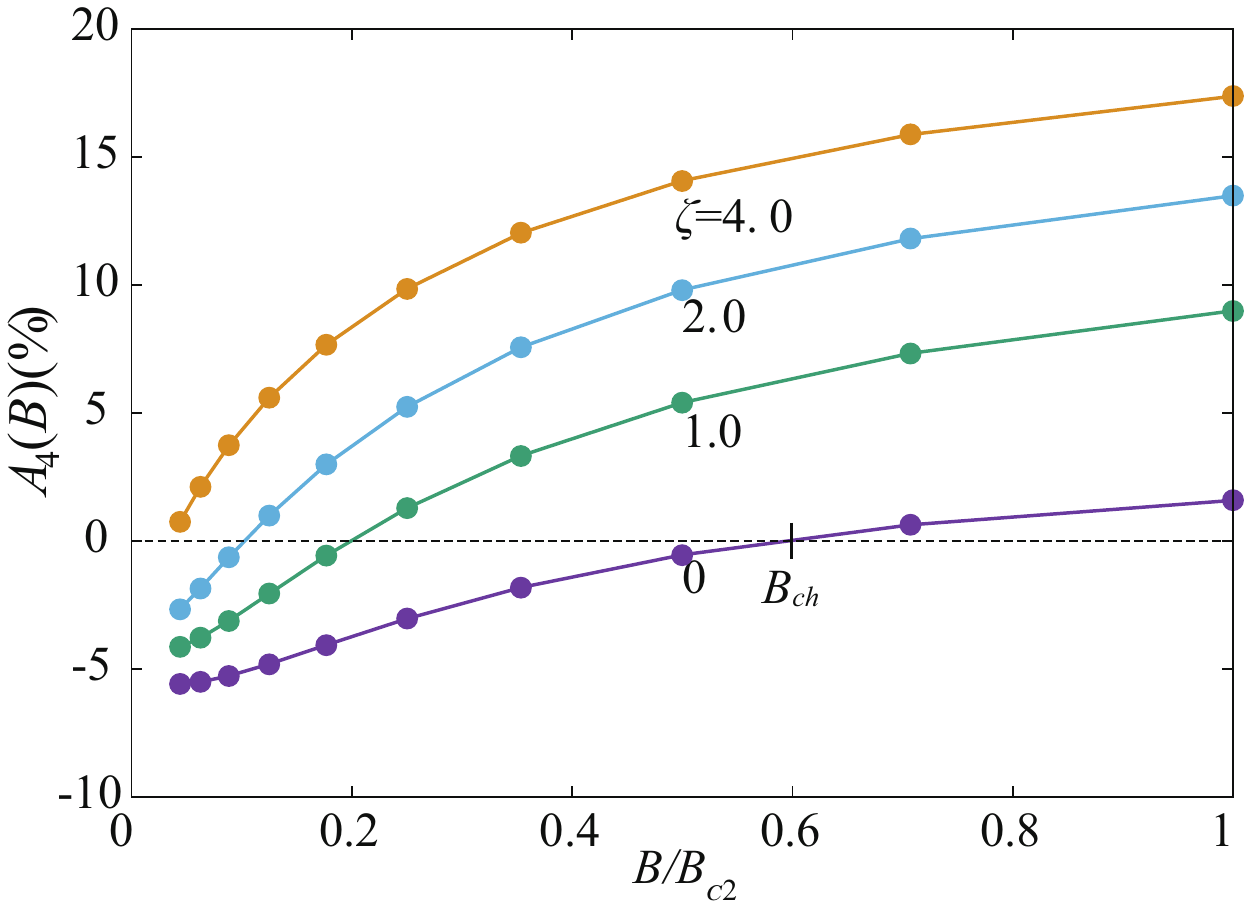}
\caption{(color online)
Field dependences of $A_4(B)$ for several $\zeta$ values in $d_{x^2-y^2}$.
The sign changing field $B_{ch}$ decreases with $\zeta$, but never disappears.
}
\label{fig3-1}
\end{figure}

\begin{figure}[tbp]
\includegraphics[width=6cm]{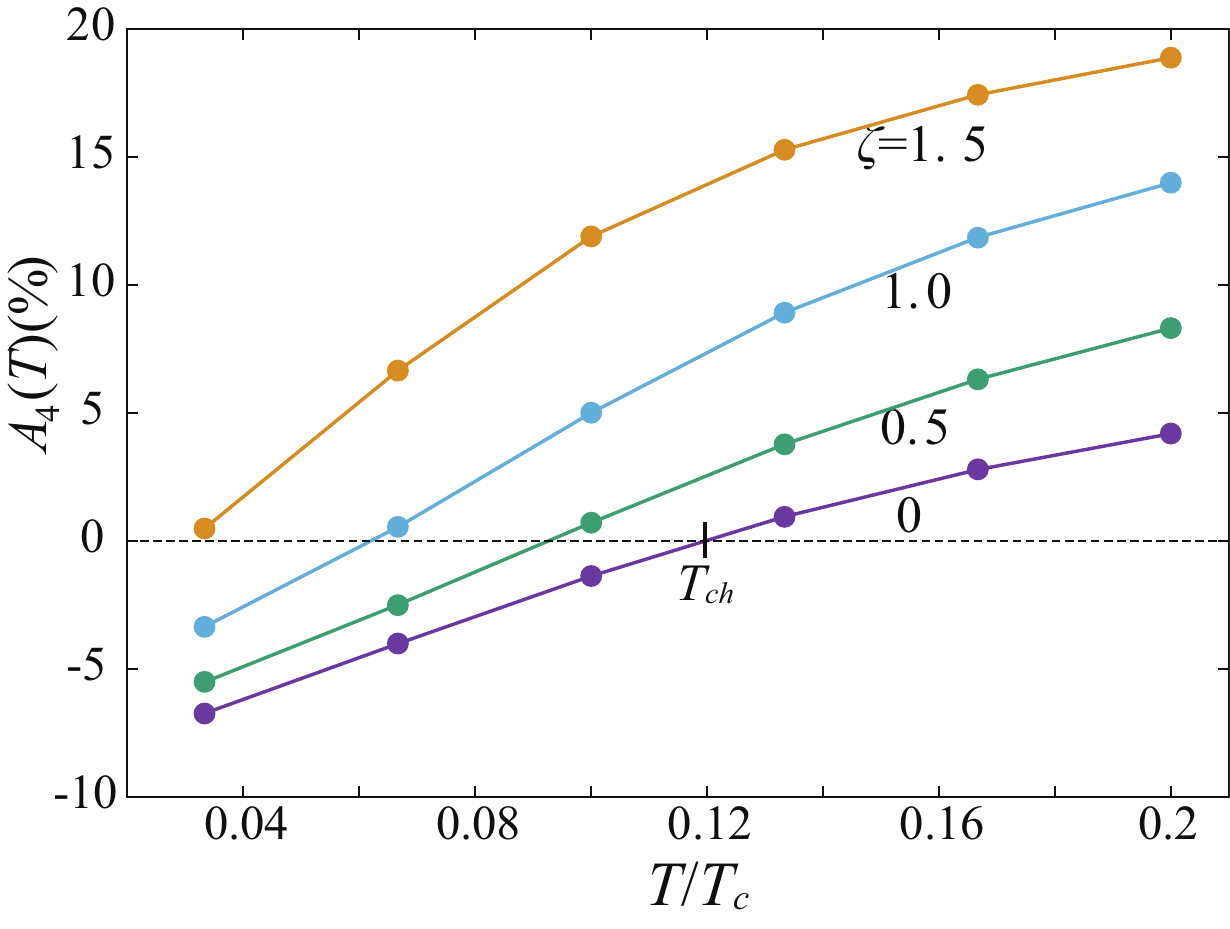}
\caption{(color online)
Temperature dependences of $A_4(T)$ for several $\zeta$ values in $d_{x^2-y^2}$.
The sign changing temperature $T_{ch}$ decreases with $\zeta$, but never disappears.
$B/B_{c2}=0.176$.
}
\label{fig3-3}
\end{figure}

\begin{figure}[tbp]
\includegraphics[width=6cm]{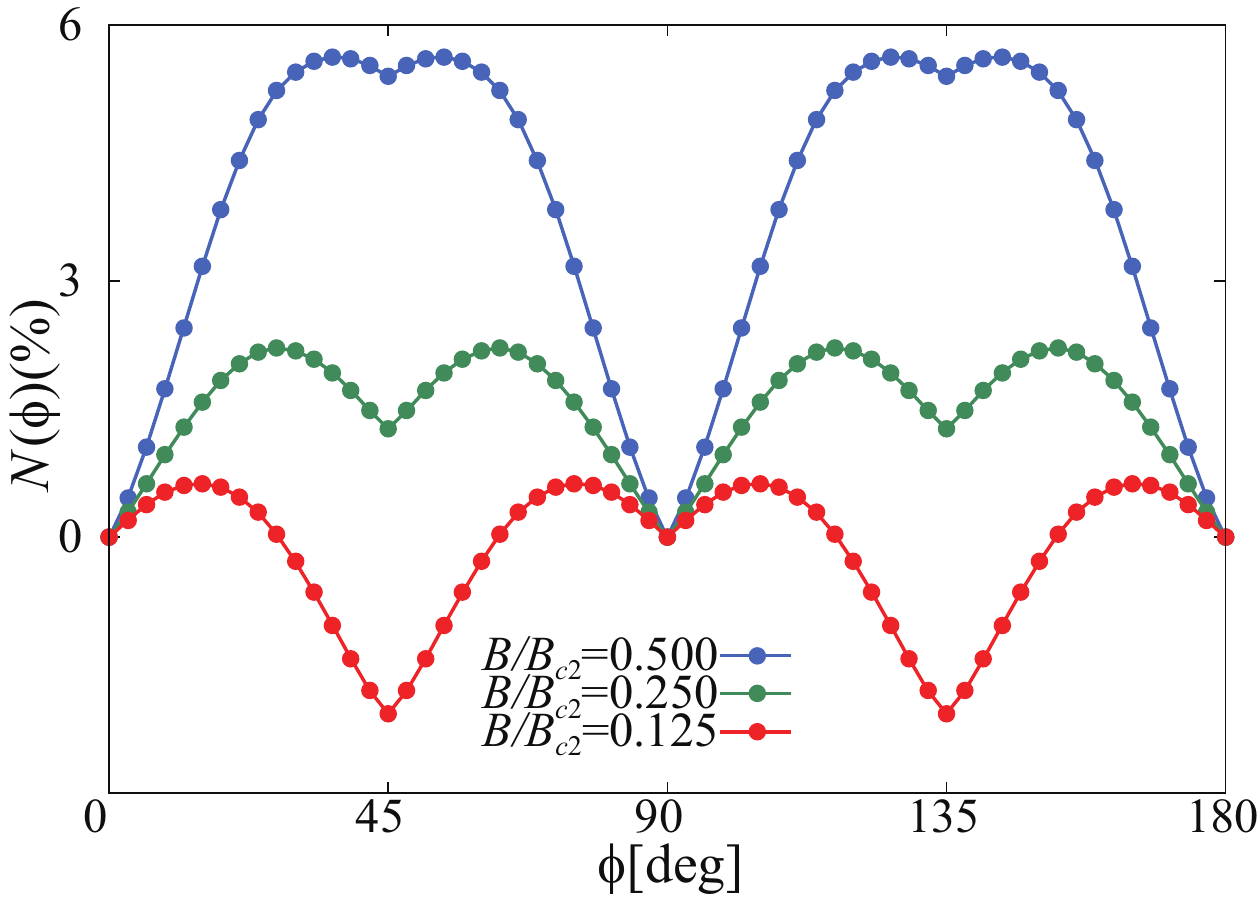}
\caption{(color online)
Oscillation patterns of $N(\phi)$ for several fields in $d_{x^2-y^2}$.
$\zeta=1.0$. It is seen that $A_4$ changes its sign under varying $B$.
 }
\label{fig3-6}
\end{figure}

As shown in Fig.~\ref{fig3-1}, for the $d_{x^2-y^2}$ symmetry case the sign changing field $B_{ch}$
in  $A_4(B)$ becomes lower as $\zeta$ increases. However, it never vanishes even for extremely larger $\zeta$
where $A_4(B)$ starts always from a negative or almost zero values near $B\sim0$.
This is also true for $A_4(T)$, as shown in Fig. ~\ref{fig3-3}.
The oscillation patterns also show a distorted periodic form far from a simple sinusoidal form
as seen from Fig.~\ref{fig3-6}.
All the above features do not agree with the experimental data~\cite{kittaka0}. Thus this is not
the case for Sr$_2$RuO$_4$.

\subsection{$d_{xy}$-symmetry}
\begin{figure}[tbp]
\includegraphics[width=6cm]{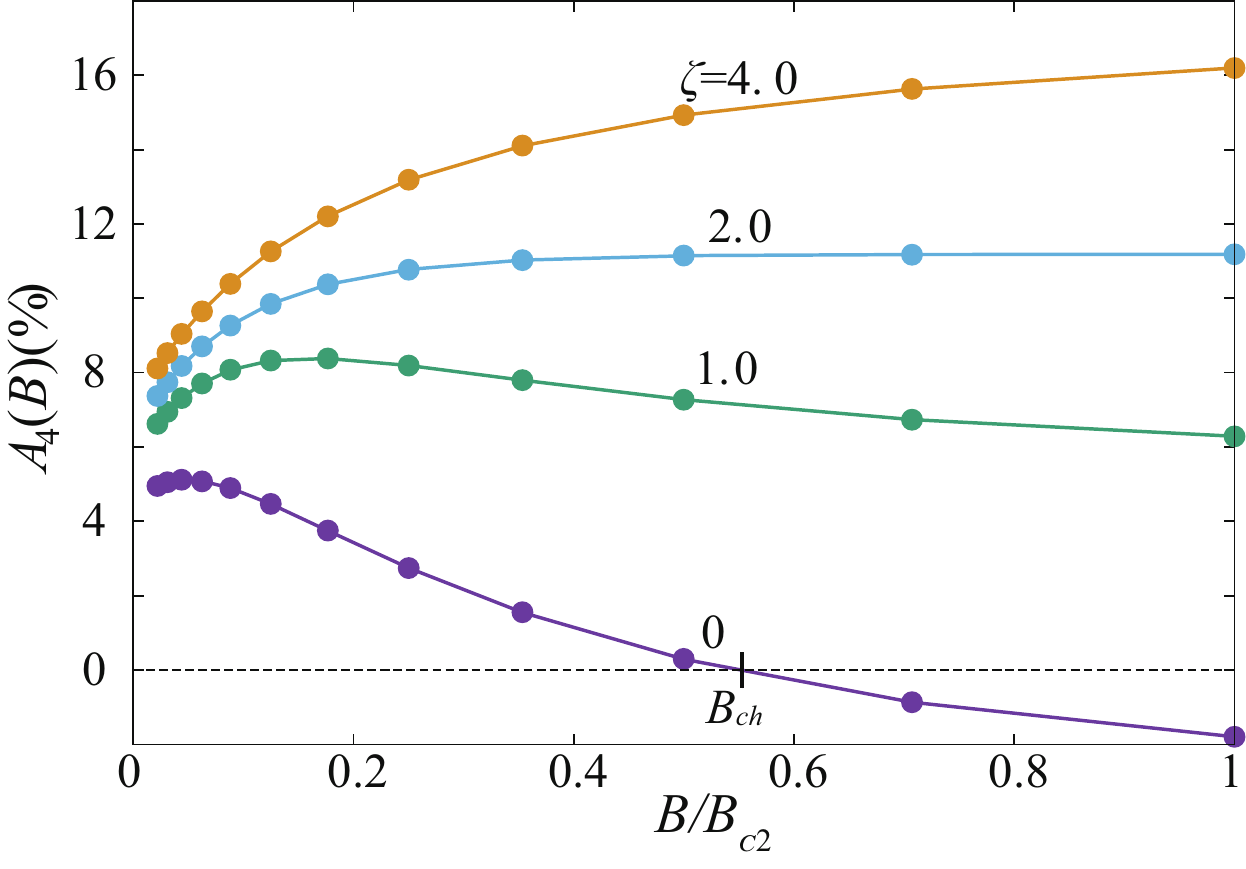}
\caption{(color online)
Field dependences of $A_4(B)$ for several $\zeta$ values in $d_{xy}$.
The sign changing field $B_{ch}$ increases with $\zeta$, and eventually disappears.
}
\label{fig3-2}
\end{figure}

\begin{figure}[tbp]
\includegraphics[width=6cm]{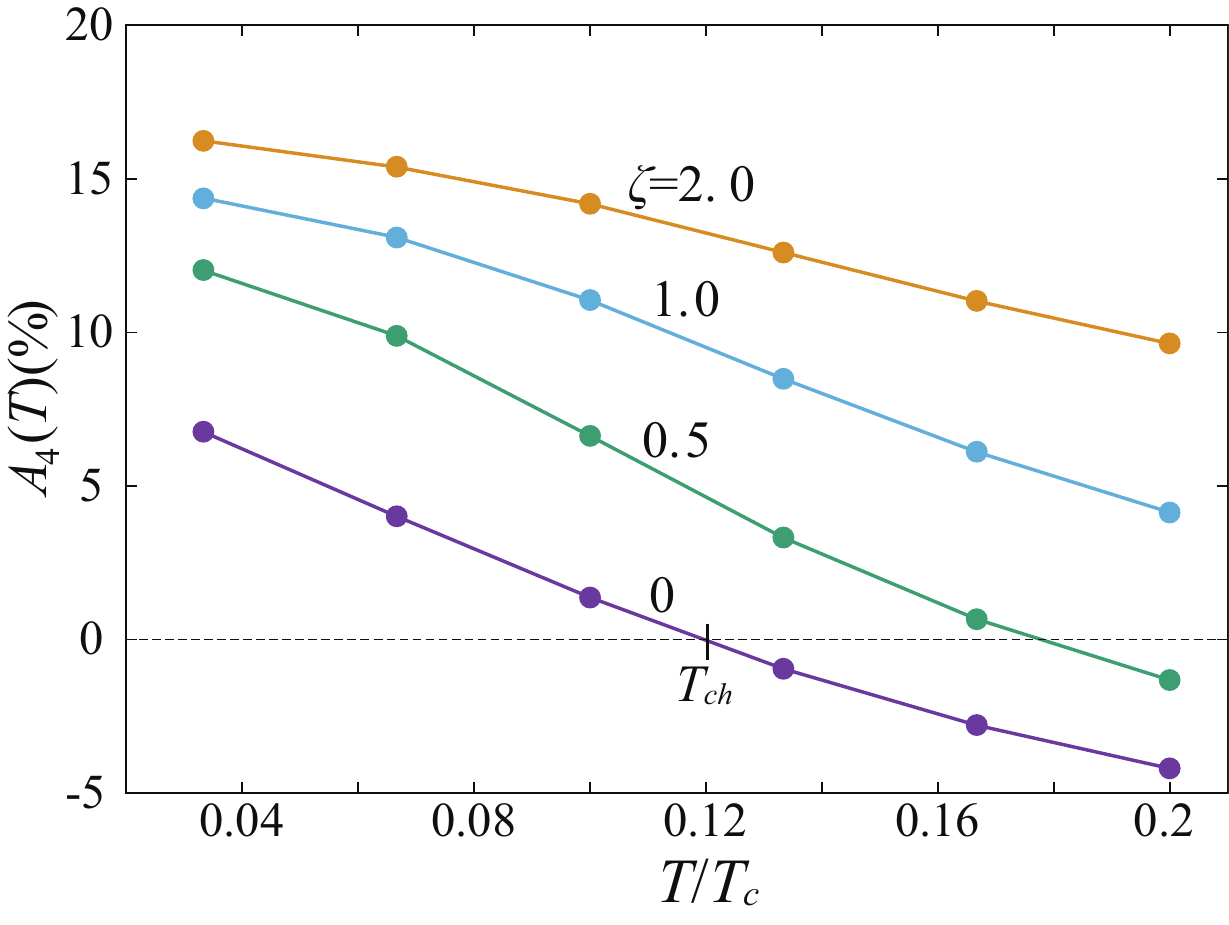}
\caption{(color online)
Temperature dependences of $A_4(T)$ for several $\zeta$ values in $d_{xy}$.
The sign changing temperature $T_{ch}$ increases with $\zeta$, and eventually disappears.
$B/B_{c2}=0.176$.
}
\label{fig3-4}
\end{figure}

\begin{figure}[tbp]
\includegraphics[width=7cm]{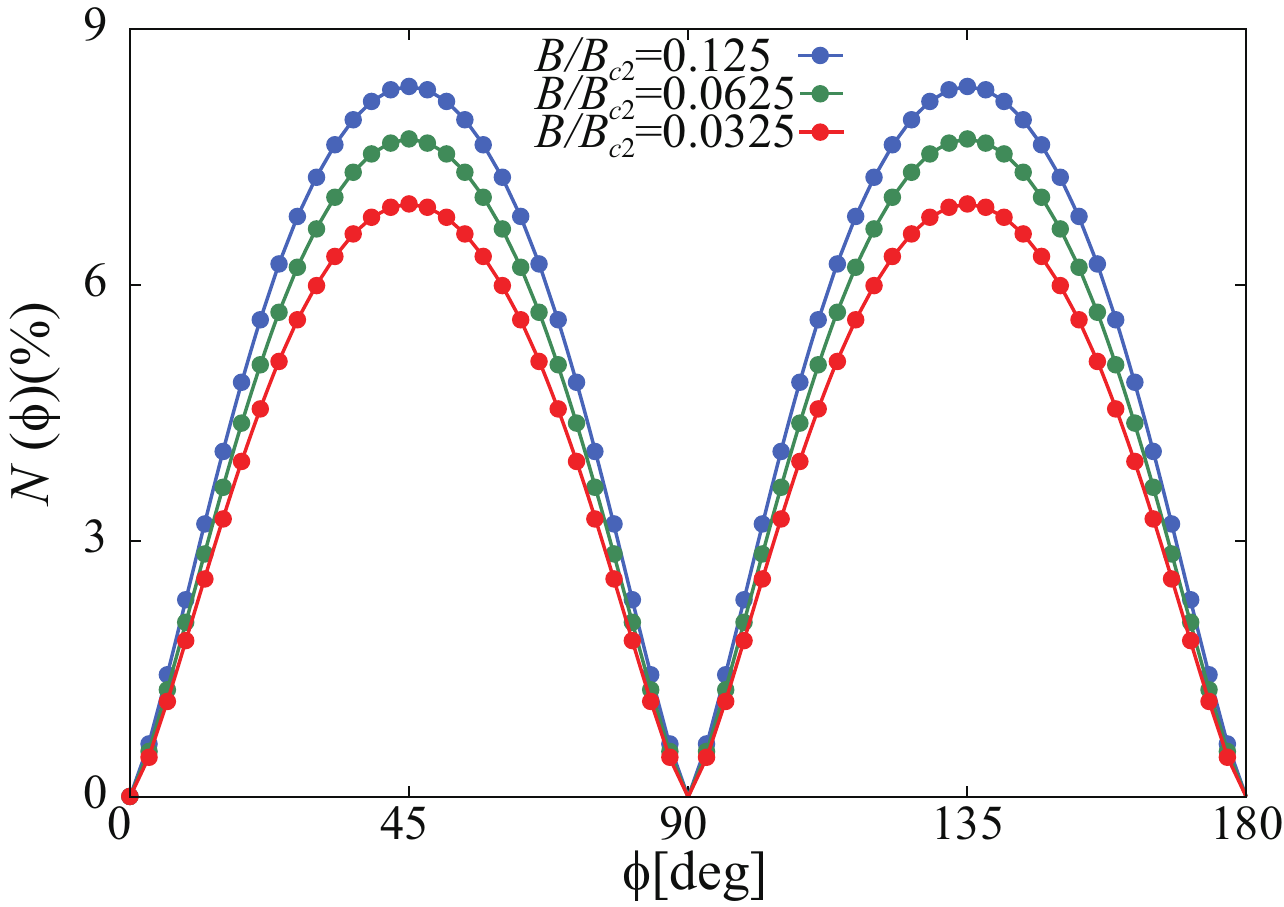}
\caption{(color online)
Oscillation patterns of $N(\phi)$ for several fields in $d_{xy}$.
$\zeta$=1.0.}
\label{fig3-5}
\end{figure}

This symmetry case seems more promising at first glance
because as seen from Fig.~\ref{fig3-2} the sign changing $B_{ch}$ in $A_4(B)$
is removed as $\zeta$ increases. Thus for a certain value of $\zeta$
the $A_4(B)$ behavior looks similar to the experimental data.
This is also true for $A_4(T)$ shown in Fig.~\ref{fig3-4}.
The sign changing temperature $T_{ch}$ tends to become higher as 
$\zeta$ increases. Therefore, $A_4(B,T)$ seems favorable for describing the data.
The oscillation patterns again  are a distorted form as seen in Fig.~\ref{fig3-5}
where we plot the results with $\zeta=1$ for selected values of $B$.
Within the accuracy of the present experiment~\cite{kittaka0}, however,  it is not possible to determine
the accurate oscillation pattern, either a simple sinusoidal or 
distorted one. Thus at this stage we cannot exclude the possibility that the vertical line nodes with 
$d_{xy}$ symmetry is realized when assuming that the $\beta$ band alone contributes to the specific heat
oscillation.
However, it is inevitable to consider the contribution for the $\gamma$ band also, which is discussed next.

\subsection{Multiband consideration}
\begin{figure}[tbp]
\includegraphics[width=8cm]{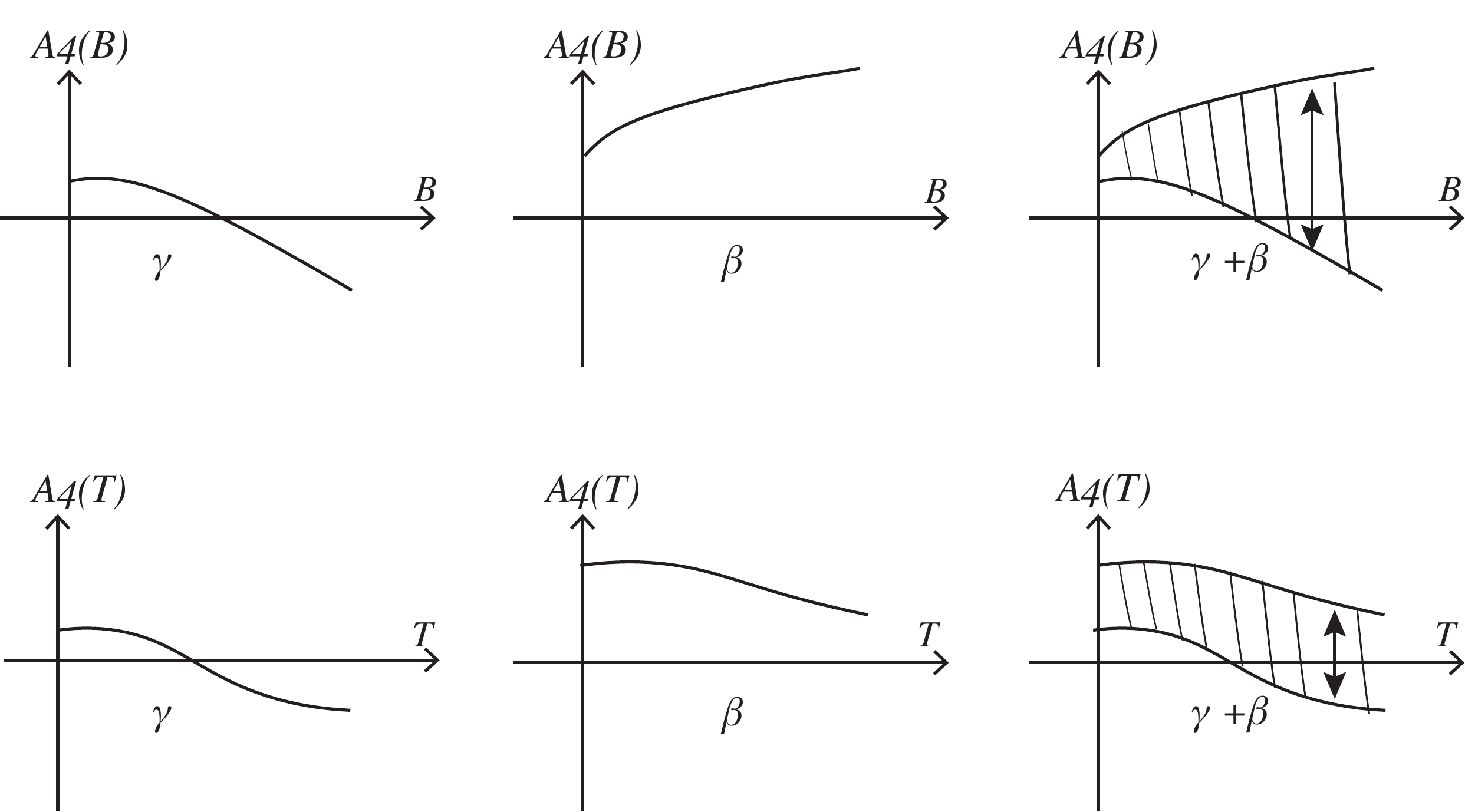}
\caption{
Possible multiband effects on $A_4(B)$ (top row)
and $A_4(T)$ (bottom row) indicated by arrows in the figures of the right column.}
\label{fig3-7}
\end{figure}

As mentioned above, the $d_{xy}$ symmetry for the $\beta$ band alone with an appropriate 
$\zeta$ value seems to explain the existing data. However, it is clear that 
$\gamma$ band with the $d_{xy}$ symmetry contributes equally to the total oscillation.
As a zero-th approximation we simply add up the two contributions by assuming that the two normal density
of states are equal (it is known that $N_{\gamma}$: 53\%, $N_{\beta}$: 37\%, and $N_{\alpha}$: 10\%
of the total DOS) and gap magnitudes $\Delta_{\beta}$= $\Delta_{\gamma}$, ignoring the $\alpha$ band for simplicity.
As shown schematically in Fig.~\ref{fig3-7}, $A_4(B)$ and $A_4(T)$ for the $\gamma$ band exhibit sign changes,
whereas those for the $\beta$ band do not. 
Thus resulting total $A_4(B)$ and $A_4(T)$
 (right column in Fig.~\ref{fig3-7}) falls somewhere in the shaded region between them.
Each may or may not exhibit the sign change, depending on other material parameters.
It may be possible to explain  the positive ``definite-ness'': $A_4(B,T)\geq0$
for nearly the entire  $B$-$T$ plane, depending on the material parameters.

This task is daunting
because there are so many adjustable microscopic parameters to tune.
For example, in order to set up the microscopic calculation for $A_4(B,T)$ using the Eilenberger equation
for the two band case, we need attractive coupling constants for the two bands $\beta$ and $\gamma$ in addition to
the Cooper pair transfer term~\cite{nakaiFF}; this includes the gap magnitude ratio $\Delta_{\beta}/\Delta_{\gamma}$, 
the Fermi velocity anisotropies for each band 
along the $c$-axis, $\Gamma_{\beta}$ and $\Gamma_{\gamma}$ which are necessary to determine $B_{c2}$ for the total system.
The in-plane Fermi velocity anisotropies, $b$ and $\zeta$ are essential.
Somewhere in the multi-dimensional parameter space there may be appropriate material parameters that explain
the positive ``definite-ness'': $A_4(B,T)\geq0$. But it is not guaranteed, so it is clear that this is quite difficult to achieve.

We conclude that the vertical line node scenario with $d_{xy}$ symmetry is not appropriate.
The other combinations, such as $d_{xy}$ on the $\beta$ band and $d_{x^2-y^2}$ on the $\gamma$ band, are
found not to be appropriate because those scenarios fail in the zero-th approximation level mentioned above.
In short, we are not denying the VLN scenario completely,
but considering the time-consuming computational burden required to solve the Eilenberger equation for the multiband case, it
is practically impossible to find a parameter set, that leads to the positive definite $A_4(B,T)\geq0$.

\section{Discussion}
\subsection{Analysis of the experimental data}
\begin{figure}[tbp]
\includegraphics[width=7cm]{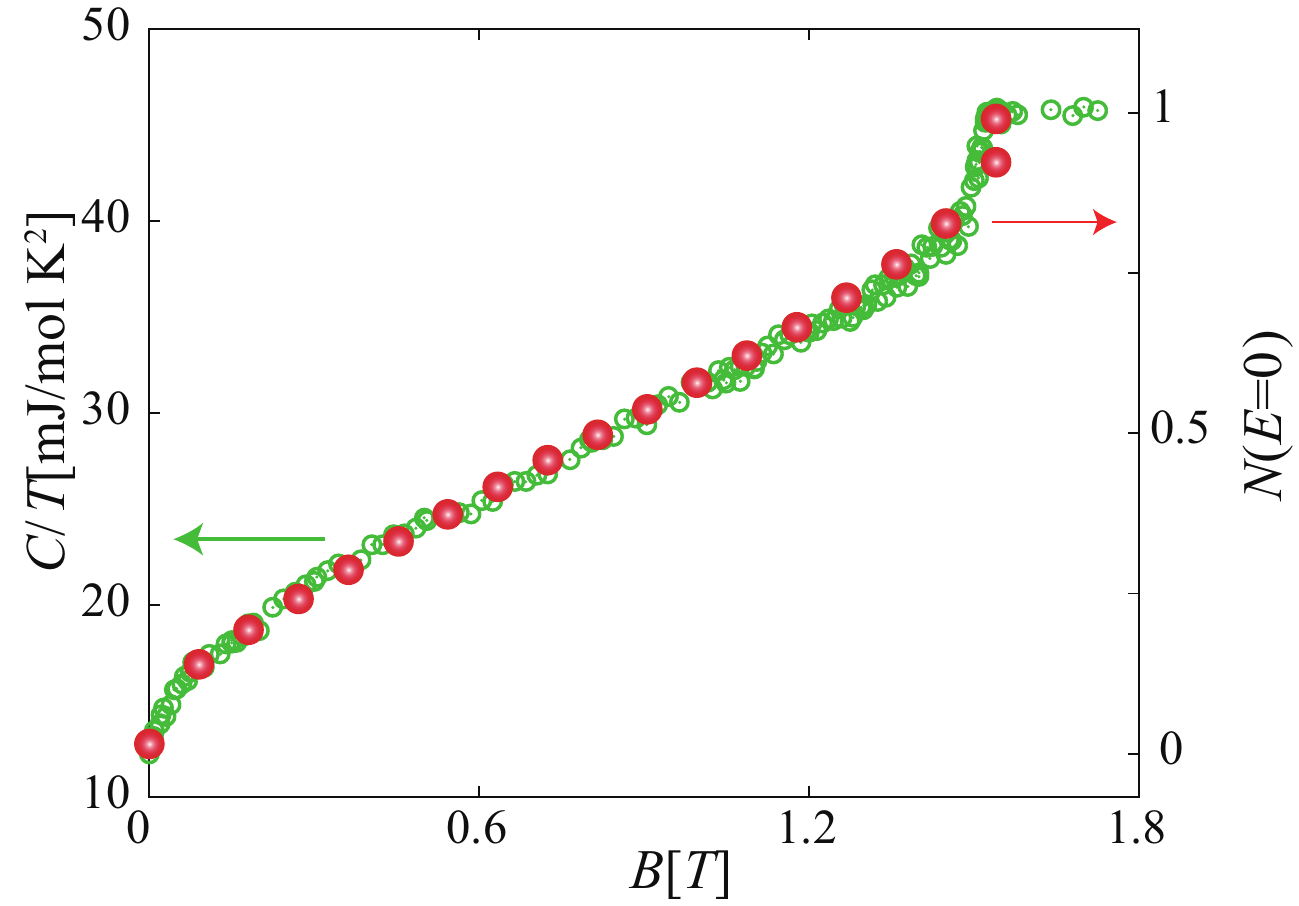}
\caption{(color online)
Comparison of calculated $N(E=0)$ shown in Fig.~\ref{fig2-14} of $\mu=0.04$
with the experimental specific heat data $C/T$ at $T=60$mK~[\onlinecite{kittaka0}].
We adjust the theoretical point at $B=0$.
}
\label{fig4-1}
\end{figure}

Having done extensive computation  for both the HLN and VLN cases,
here we discuss the implications of our results and analyze the experimental data,
which are summarized by the four items [1]-[4] mentioned in the Introduction.
Before that, we point out the importance of PPE in analyzing the data.
To demonstrate this, we compare the theoretical data for the field evolution of
ZDOS $N(E=0)$ under PPE (see Fig.~\ref{fig2-14}) and the experimental data in Fig.~\ref{fig4-1}.
At lower $B$, $C/T$ increases quickly, reflecting the nodal quasi-particles reminiscent of the Volovik $\sqrt B$.
Then $C/T$ slowly rises in the middle $B$, and finally it exhibits a jump associated with a first order transition at $B_{c2}$
due to PPE. These features are captured by our theoretical results.
Almost perfect agreement between the theoretical and experimental results 
implies that PPE is inevitable for the following analyses.

(I) Absence of the sign change in $A_4(B, T)$
If the $\gamma$-band  which is well approximated by the $b$-model has VLN
and the major band, namely $\Delta_{\gamma}>\Delta_{\beta}$,
$A_4(B, T)$ should exhibit the sign change   along both $B$ and $T$ axes
because $N_{\gamma}=53\%$ is the largest and dominates the oscillation.
However, those conditions are not met, the $\beta$-band which is modeled by the $\zeta$-model
plays a role in determining $A_4(B, T)$.
When  $\zeta$ is large enough, $A_4(B, T)$ may  exhibit no sign change under the
assumption that the $\beta$-band alone dominates the oscillation.
However, this is unlikely because of $N_{\beta}=37\%$ and $\Delta_{\gamma}\sim\Delta_{\beta}$
at most, the ratio of which is not known precisely.
We assigned $\Delta_{\beta}=\Delta_{\gamma}/2$ in our previous paper\cite{nakaiFF} by analyzing SANS experiments\cite{morten1,morten2}.
Thus we consider the multiband effect when solving the Eilenberger equation
for two bands or three bands. As already mentioned, it is a daunting task to achieve.
The educated guess is that the ``positive definite-ness''  of $A_4(B, T)$ is virtually impossible to reproduce 
in terms of VLN considering  the delicate balance of the $A_4^{\gamma}$ and $A_4^{\beta}$ contributions.

(I\hspace{-.1em}I) $A_4(B)$ behavior According to the microscopic Eilenberger calculation for the $b$-model, 
$A_4(B)$ starts at a finite value at lower $B$ and increases with $B$, 
reaching a maximum at $B\simeq 0.3B_{c2}$ (see Fig.~\ref{fig2-8}). 
$A_4(B)$ smoothly decreases almost linearly toward $B_{c2}$ 
where $A_4(B_{c2})=0$ precisely.

\begin{figure}[tbp]
\includegraphics[width=7cm]{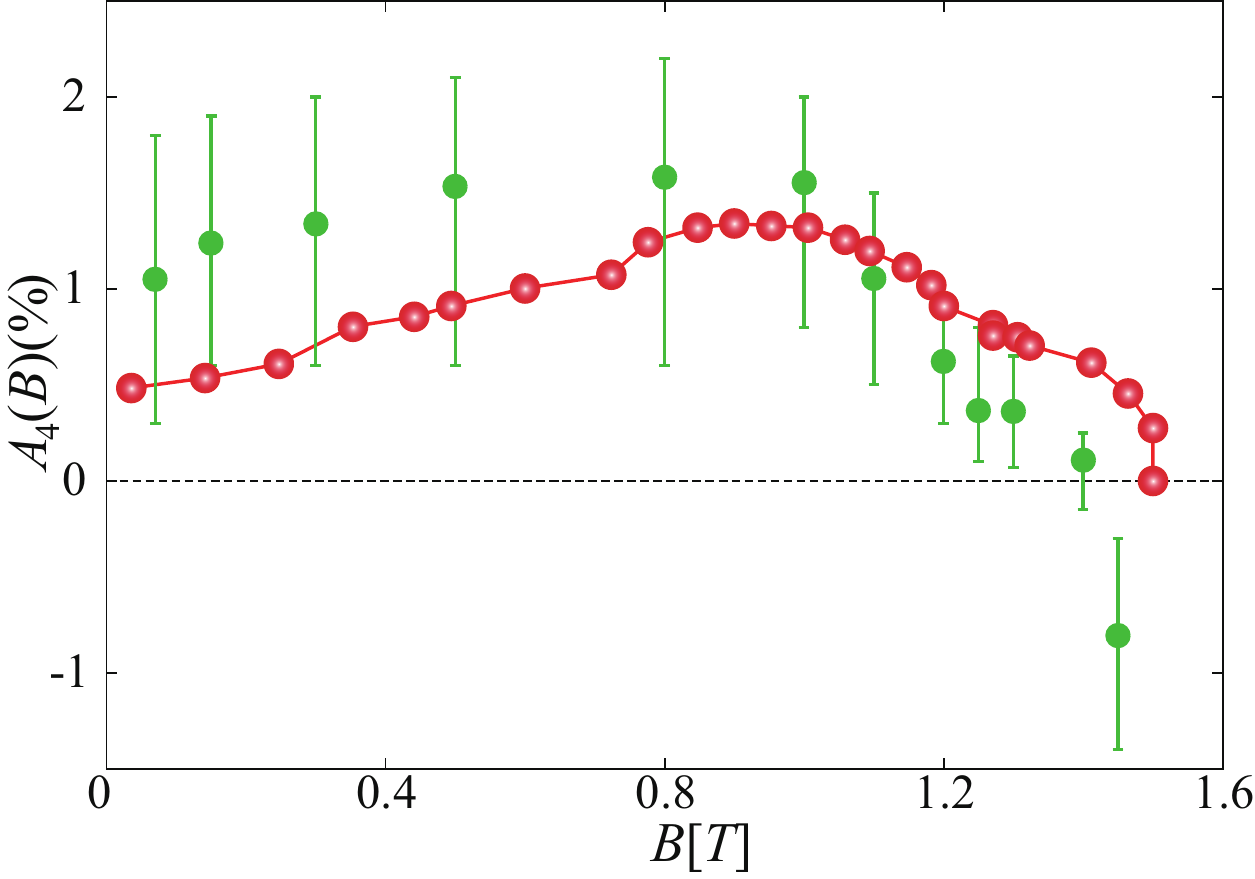}
\caption{(color online)
Comparison of calculated $A_4(B)$ shown in Fig.~\ref{fig2-11} with the experimental data
at $T=100$mK~[\onlinecite{kittaka0}]. We show the theoretical fit curve by choosing $B_{c2}=1.5$T where the vertical scale is arbitrary.
}
\label{fig4-2-1}
\end{figure}

\begin{figure}[tbp]
\includegraphics[width=7cm]{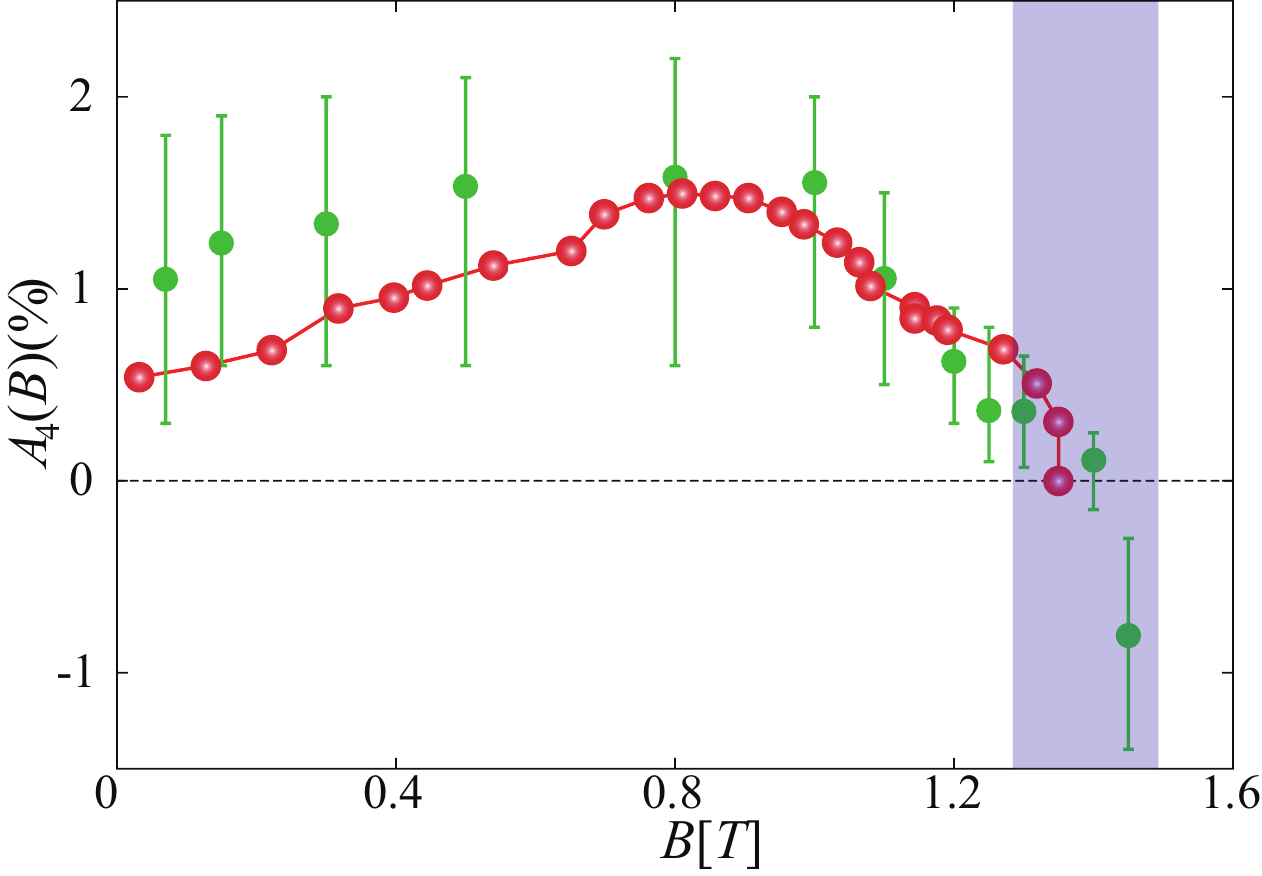}
\caption{(color online)
Comparison of calculated $A_4(B)$ shown in Fig.~\ref{fig2-11} with the experimental data
at $T=100$mK~[\onlinecite{kittaka0}]. We show the theoretical fit curve by choosing $B_{c2}=1.35$T  where the vertical scale
is arbitrary.
The agreement is far better for this choice.
We highlight the anomalous field region near B$_{c2}$ by a color bar.}
\label{fig4-2-2}
\end{figure}

By using the DOS scaling we obtain $A_4(B)$ under PPE
which explains well the experimental data as demonstrated in Figs. \ref{fig4-2-1} and \ref{fig4-2-2}.
We notice the followings:
\renewcommand{\labelenumi}{(\Alph{enumi})}
\begin{enumerate}
\item The obtained $B_\textrm{max}/B_{c2}=0.7$ is achieved only by taking PPE into account as shown in  Fig.~\ref{fig2-11}.
Thus PPE is essential in understand the physics of Sr$_2$RuO$_4$, otherwise it is at $B_\textrm{max}/B_{c2}\sim 0.3$.
\item As seen from Fig.~\ref{fig4-2-2}, almost perfect fitting is achieved by shifting $B_{c2}$  so as to coincide $B$ with the field at 
$A_4(B)=0$ where the ``theoretical $B_{c2}$'' is situated.
This means that the actual $B_{c2}$ is enhanced.
\item The observed $A_4(B)\leq 0$ region colored in  Fig.~\ref{fig4-2-2} appears above this field and corresponds to the ``enhanced'' region. 
This anomalous ``enhanced''  field region corresponds to the FFLO phase expected for a clean superconductor with
strong PPE, a condition that is indeed fulfilled in the present Sr$_2$RuO$_4$ known as a super-clean 
system. The mean free path is 140nm -- 300nm.~\cite{mackenzie}
The precise identification of the discovered region calls for further investigation both experimentally
and theoretically.
\end{enumerate}

(I\hspace{-.1em}I\hspace{-.1em}I) Narrow $T$ region for finite $A_4(T)$ 
According to the Doppler shift picture, which is shown schematically in Fig.~\ref{fig2-6}, the characteristic 
energy window $E_D$ by the Doppler shift is confined 
in a finite narrow energy region around $E=0$.
According to our numerics $A_4(T)$ calculated by Eq. \eqref{A4T} yields a finite value up to at most $\sim 0.2T_c$.
This contrasts with that of the VLN cases;
The angle dependent DOS change $\delta N(E)$~\cite{vekhter}
which drives the oscillation persists at a much higher  energy, thus 
leading to the wider $T$-region of $A_4(T)$~\cite{hiragi}. 
Therefore this experimentally demonstrated narrow $T$ region is an eminent characteristic of HLN.

\begin{figure}[tbp]
\includegraphics[width=8cm]{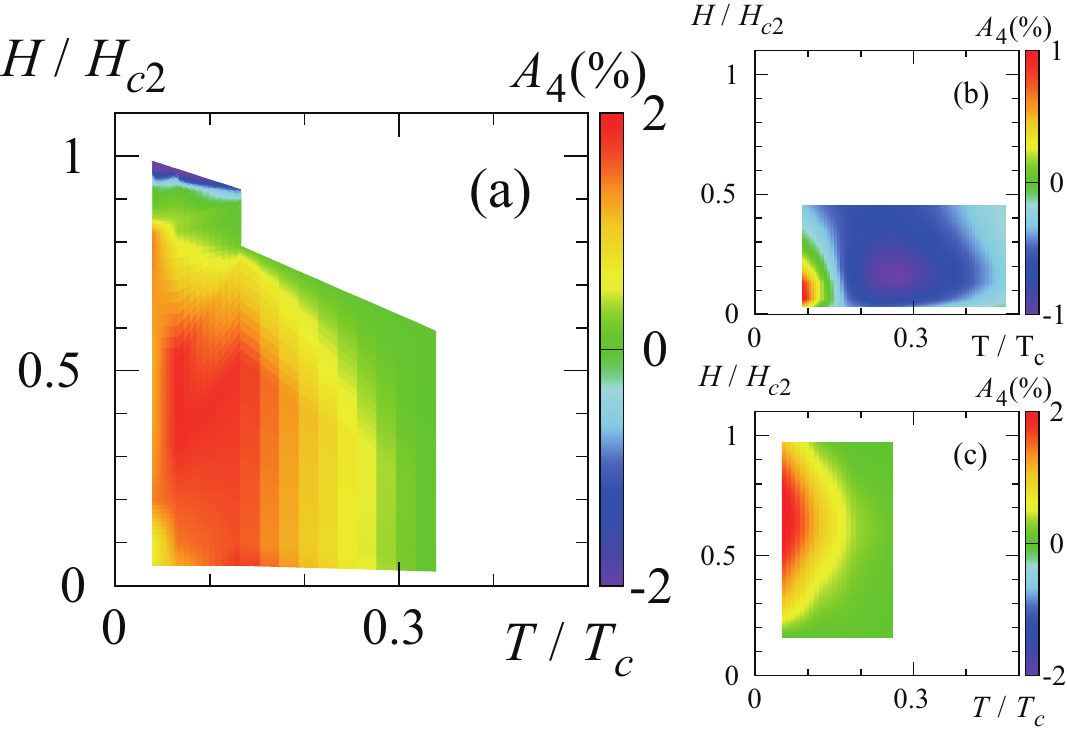}
\caption{(color online)
Comparison of landscapes of (a) $A_4(B,T)$ determined experimentally~[\onlinecite{kittaka0}],
(b) vertical line nodes: $d_{xy}$ calculated for $b=0$ and $\mu=0$~[\onlinecite{hiragi}], and (c) horizontal line nodes for $b=0.5$ and $\mu=0.04$.
}
\label{fig2-13-2}
\end{figure}

(I\hspace{-.1em}V) Simple landscape of $A_4(B,T)$
The experimental landscape of $A_4(B,T)$ is quite simple (see Fig.~\ref{fig2-13-2} (a)).
Most of the $B$-$T$ plane is covered by $A_4(B,T)\geq0$ except for just
below the $B_{c2}$ region with $A_4<0$.
This landscape is well reproduced by HLN shown in Fig.~\ref{fig2-13-2} (c).
This is contrasted with the typical VLN case is shown in Fig.~\ref{fig2-13-2} (b)~\cite{hiragi}
where a rather complicated landscape with a local maximum, local minimum and valley
form the landscape.
Thus it is clear that HLN is superior to VLN in this point of view.

\subsection{Unified picture of Sr$_2$RuO$_4$ and future prospects}
Having discussed the four items of the experimental findings in light of the
present theory and concluding that the realized gap structure is 
described by horizontal line nodes, we are now in a position to describe the overall superconducting
properties of Sr$_2$RuO$_4$ from a unified viewpoint.

In the group theory classified pairing symmetries within the chiral $p$-wave there is no state with horizontal line nodes\cite{sigrist,ozaki1,ozaki2} except for  $(k_x+ik_y)\cos k_z$ that has accidental nodes~\cite{hasegawa}.
The overall pairing symmetry could be consistent with $d_{3k_z^2-1}$ and the chiral d-wave form $(k_x+ik_y)k_z$
or $(k_x+ik_y)\cos k_z$.
The latter two are time-reversal symmetry broken, thus those are
compatible with $\mu$SR\cite{luke} and Kerr rotation\cite{kapitulnik} experiments which claim it.
In order to distinguish those states, we propose carrying out a 
 spin gap and/or spin resonance experiment by inelastic neutron scattering at 
$Q_\textrm{res}=(1/3,1/3,0.15 (=0.85))$ or (1/3,1/3,0.35 (=0.65)) for the former 
and $Q_\textrm{res}=(1/3,1/3,0.5)$ for the latter two  in the reciprocal units.
Since those distinctive different reciprocal space points can be probed by neutron scattering experiment in principle.


The analysis of $A_4(B)$ mentioned in Item [2] strongly suggests that the extra state appears above the nominal ``$B_{c2}$''.
The two characteristics of the enhanced $B_{c2}$ and the extra state may correspond to the FFLO.
In fact $A_4(B)<0$ just characterizes this high field phase.
The in-plane anisotropy $B_{c2}(\phi)$ is consistent with $A_4(B)<0$,
namely, $B_{c2}(\phi=45^{\circ})>B_{c2}(\phi=0^{\circ})$ means $N(E=0,\phi=45^{\circ})<N(E=0,\phi=0^{\circ})$.
According to Kittaka \textit{et al.} \cite{kittakaBc2} the $B_{c2}(\phi)$ anisotropy appears below $T<T_\textrm{1st}=0.8$K and above $B>B_\textrm{1st}=1.2$T
which coincides with the first order transition line along the $B_{c2}$ line.
Below this point $B<B_\textrm{1st}$ the $B_{c2}$ line is ordinary second order
and the $B_{c2}(\phi)$ anisotropy disappears simultaneously.
This phase diagram in the $B$-$T$ plane is expected for the FFLO,
namely $T_\textrm{1st}/T_c=0.8K/1.5K=0.53$ is very near the ideal triple point, i.e. the so-called Lifshitz point
$T_{tri}/T_c=0.56$ below which  the FFLO~\cite{nakanishi,suzukikenta} starts to appear.

Concerning the question regarding which band plays the major role for superconductivity 
among $\alpha$-, $\beta$- and $\gamma$-band, we consider that the $\beta$-band has a
larger gap than the $\gamma$-band, namely $\Delta_{\beta}>\Delta_{\gamma}$
because the observed $\Gamma_{VL}\sim 60$ just corresponds to $\Gamma_{\beta}\sim 60$
rather than $\Gamma_{\gamma}\sim 180$ at least near $B_{c2}$.
However, at first sight it is at odds with the absence of the in-plane $B_{c2}(\phi)$ anisotropy above
$T>T_\textrm{1st}=0.8$K when one considers the anisotropic square-like Fermi surface shape in the $ab$ plane
that we model as the $\zeta$-model. This easily gives  rise to the $B_{c2}^{\beta}(\phi)$ anisotropy
if $\zeta$ is large (when $\zeta$=2, $B_{c2}(\phi)$ anisotropy defined by $B_{c2}(\phi=0^{\circ})$/$B_{c2}(\phi=45^{\circ})$ is $\sim$1.13).
It should be noticed that in the $b$-model for the $\gamma$-band the in-plane anisotropy 
$B_{c2}^{\gamma}(\phi)$ is absent irrespective of the $b$ value as mentioned before in Eq.(\ref{unity}).
This paradox may be solved by either assuming that $\zeta$ may not be so large
or that a substantial in-plane gap anisotropy $\Delta_{\beta}(\phi)$ is present
that cancels the Fermi surface anisotropy modeled by  $\zeta$.
Here we prefer the former scenario because the latter would require a large additional condensation penalty.
After all, the $\zeta$ value for the $\beta$ band may not be so large.
This is currently an open question.

In view of the recent remarkable series of uni-axial stress experiments, which
reported the Knight shift change~\cite{brown} below $T_{\rm c}$ and continuity of $T_{\rm c}$
under varying uni-axial stresses without cusp features~\cite{hicks1,hicks2,hicks3} expected for degenerate 
representations such as $p_x+ip_y$ or $d+id$,
it is natural to consider that Sr$_2$RuO$_4$ is a spin-singlet superconductor.
If we pick up the appropriate pairing state among the $d$-wave category
$d_{3k_z^2-1}$ symmetry with off-symmetry horizontal line nodes is the most
viable choice, which is consistent with the present experiment\cite{kittaka0} and theoretical
analysis.
Other gap symmetry with accidental nodes may be present.
More investigation is required to finally identify the pairing symmetry in this system.

\textit{Note added in proof.} Quite recently, Iida, \textit{et al.} \cite{iida} have observed a spin gap at ($1/3, 1/3, 1/2$) in the reciprocal lattice units, which is indeed fully consistent with the horizontal line nodes in Sr$_2$RuO$_4$.

\begin{acknowledgments}
We are thankful for close collaboration  with the experimental group of T. Sakakibara, S. Kittaka, and N. Kikugawa.
This work was supported by  JSPS KAKENHI Grant Numbers 17K05553 and 15K17715.
A part of the numerical calculations was performed by using the HOKUSAI supercomputer system in RIKEN.
\end{acknowledgments}

\appendix
\section{Full gap case}
\begin{figure}[tbp]
\includegraphics[width=7cm]{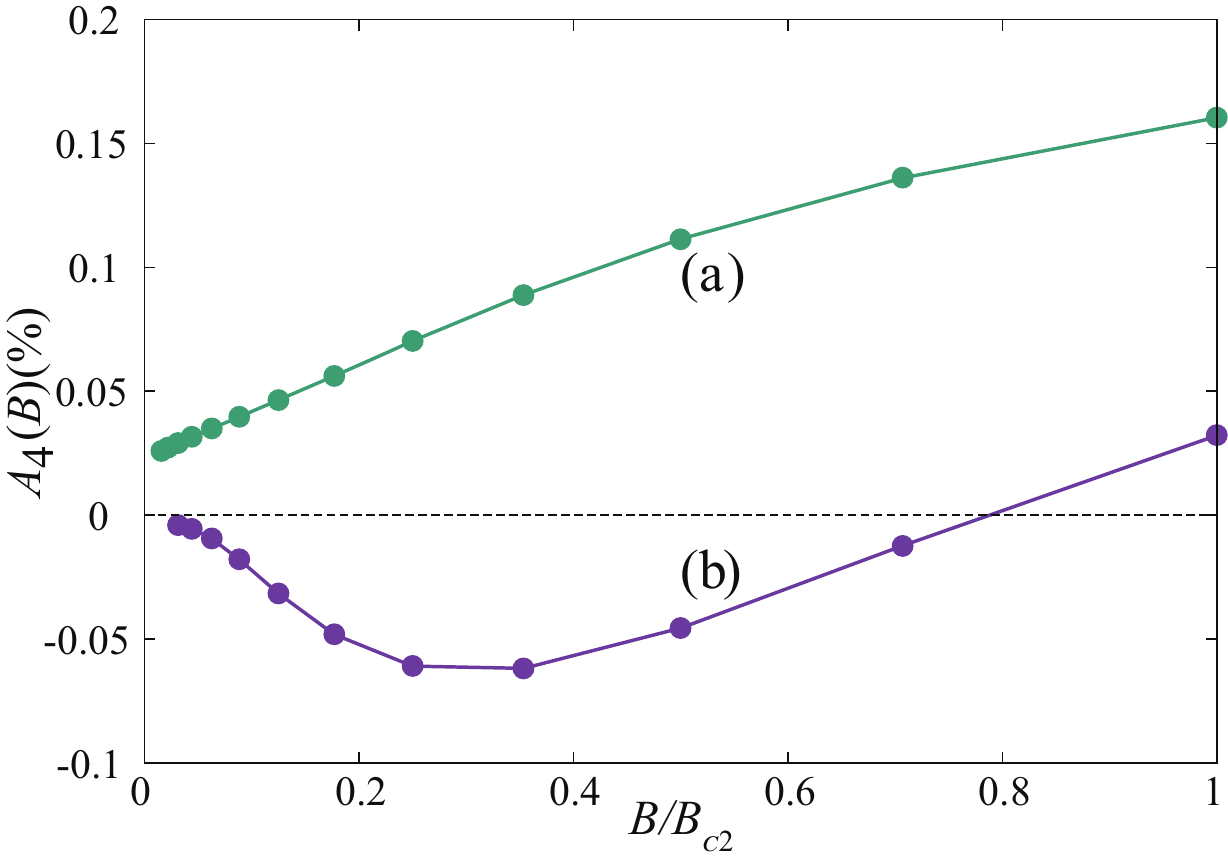}
\caption{(color online)
Field dependences of $A_4(B)$ for 
HLN (a) and full gap (b) as calculated by KPA. $b=0.33$. $A_4(B)>0$ for all $B$ and remains finite when $B\rightarrow0$
for the HLN case (b) while in the full gap case (a) $A_4(B)\rightarrow0$ as $B\rightarrow0$ and changes its sign as $B$ increases.}
\label{fig2-1}
\end{figure}

\begin{figure}[htbp]
\includegraphics[width=8cm]{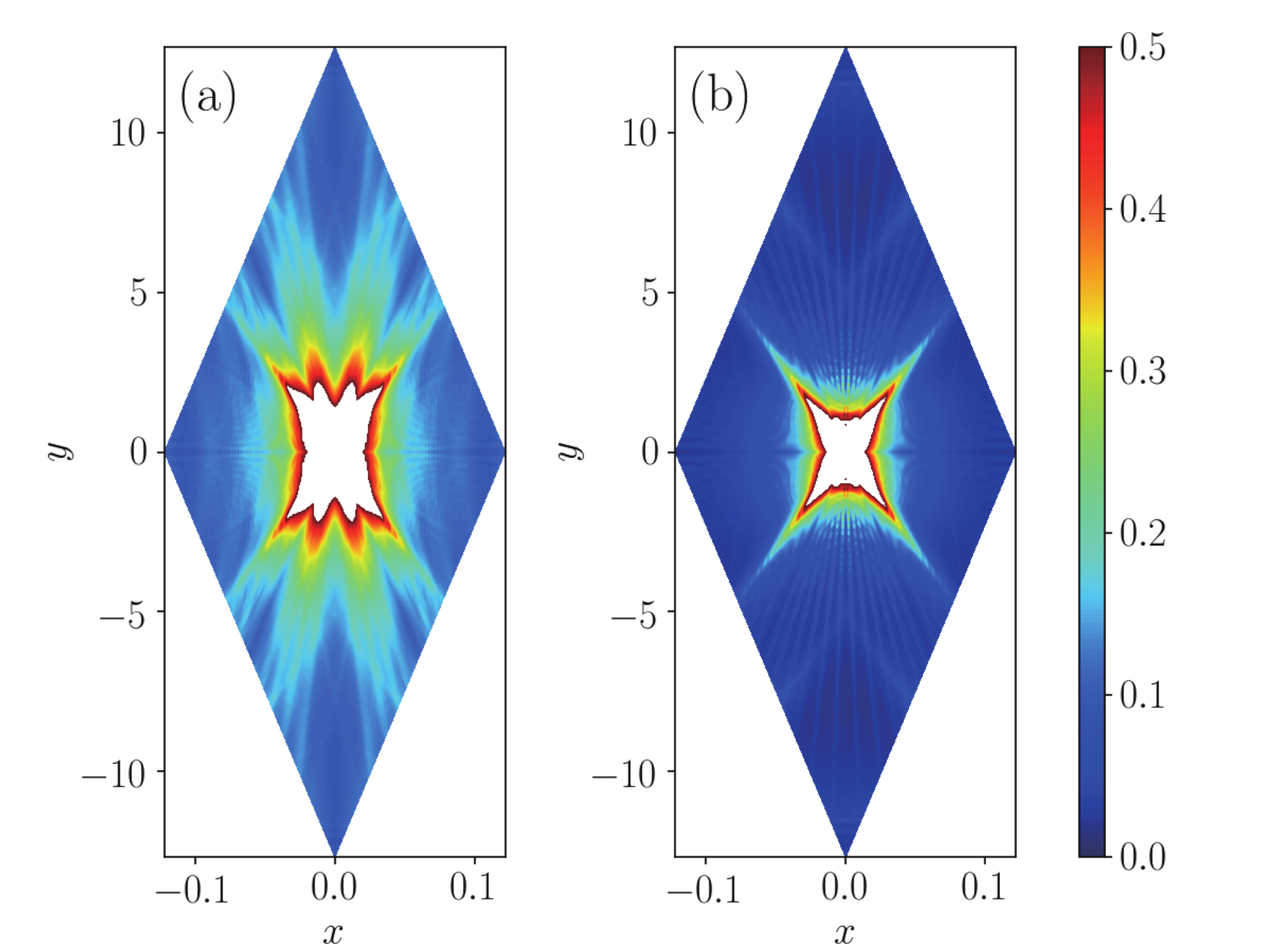}
\caption{(color online)
Landscapes of the zero energy DOS for two gap functions,
horizontal line nodes (a) and full gap (b) calculated by using full Eilenberger theory
at the same field B=2; the results are shown within one unit cell with a vortex core at the center. 
ZDOS is extended for the HLN case ($N(E=0)=0.23$) while it is concentrated 
and localized at the vortex core for a full gap ($N(E=0)=0.10$).
Note that the unit cell is distorted due to the anisotropy $\Gamma=60$.}
\label{ZDOS}
\end{figure}

\begin{figure}[tbp]
\includegraphics[width=7cm]{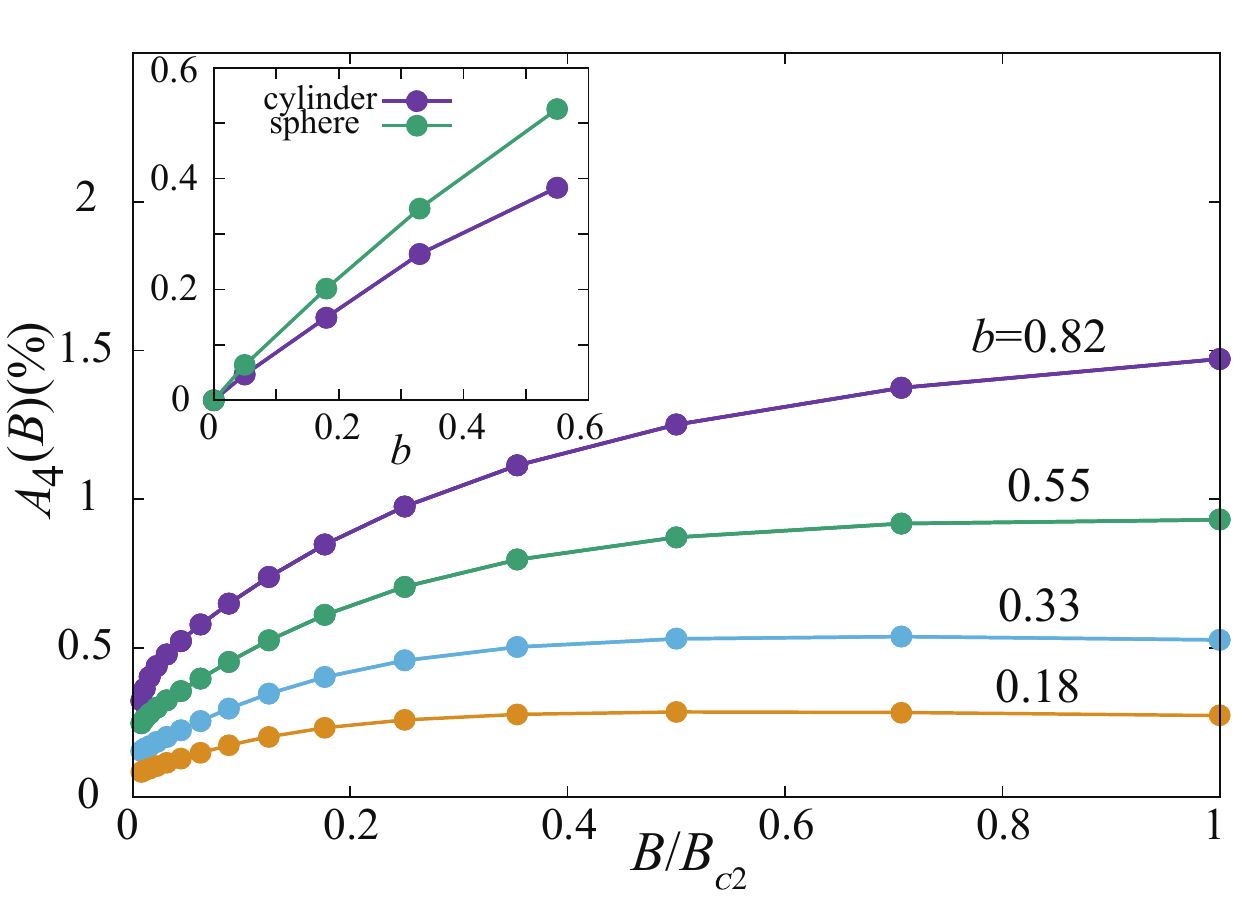}
\caption{(color online)
$A_4(B)$ for various $b$ values for the 3D Fermi sphere case,
showing $\sqrt B$-like increases first and then nearly
saturating to a constant as $B\rightarrow B_{c2}$.
The inset shows $A_4(B)$ as a function of $b$, demonstrating that
$A_4(B)\propto b$ for both 2D cylinder and 3D sphere cases at $B/B_{c2}=0.17$.
}
\label{fig2-3}
\end{figure}

It is instructive to see the full gap case compared with HLN case shown in the main text.
As seen from Fig.~\ref{fig2-1} where $A_4(B)$ is displayed for  HLN (a) and full gap (b) cases
with KPA,  we see the followings for the full gap case (b):

\noindent 
(1)  $A_4(B)\rightarrow0$ as $B\rightarrow0$,

 \noindent
(2) for lower field $B$, $A_4(B)<0$, and 

 \noindent
(3) after reaching a minimum $A_4(B)$ changes its sign.
These results are contrasted with the case of HLN, namely
$A_4(B) >0$ always positive, it monotonically increases, and it approaches
 a finite value as $B\rightarrow0$.
 
The results for a full gap case agree with those reached by full self-consistent 
Eilenberger calculation (see Fig.~2 in Ref. \onlinecite{miranovic1}).
The differences in the tendency of $A_4(B)$ as $B\rightarrow0$ for nodes and a full gap cases 
an important signature of the gap structure that appeared as we examined 
the experimental data.

The contrasting sign difference in $A_4(B)$ for HLN and full gap cases in lower fields is understood as follows:
In the full gap case, the angle-resolved zero energy DOS (ZDOS) $N({\phi})$ reaches a maximum in the $\phi=0$ direction
since $N({\phi})\propto 1/v_F(\phi)$ while in HLN $N({\phi})\propto v_F(\phi)$ due to the Doppler shift.

This difference originates from the fundamentally different nature of quasi-particles with zero energy:
As seen from  Fig.~\ref{ZDOS} we compare the landscapes in a vortex lattice unit cell for two cases at the same field.
The zero-energy quasi-particles are extended in HLN (a) while they are localized and confined in the vortex core region in the
full gap case (b).
Therefore, in the former they fully participate in the superfluid screening current velocity ${\bf v}_s$ around the vortex core.
In the main text, we focus on those extended
nodal and also core-localized quasi-particles with zero-energy associated with HLN, 
which play a fundamental role in the specific heat
oscillations.

\section{3D Fermi sphere case}
We show the KPA results of $A_4(B)$ for HLN for  three-dimensional (3D) spherical Fermi surface in Fig.~\ref{fig2-3}.
It is seen from this that $A_4(B)$ in 3D nearly saturates for higher fields, and that it increases rather quickly.
By increasing the Fermi velocity anisotropy $b$ that is ntroduced in Eq.~(\ref{vF}) the amplitude $A_4$ grows.
The growing rate is linear in $b$ at least for smaller and moderate $b$ values as seen from 
the inset of Fig.~\ref{fig2-3},
where $A_4(B)$ is plotted under a fixed $B$ as a function of $b$ for both 2D and 3D cases.
$A_4(B)$ tends to nearly saturate or slowly increase at higher fields for the 3D case because 
the DOS is given by
\begin{eqnarray}
N(E)={\pi\over2}{|E|\over \Delta_0}  \ \ \ \ \ \ (|E|<\Delta_0)
\end{eqnarray} 
all the way up to the gap edge\cite{sigrist}, namely the slope of the DOS: $dN(E)/dE$ does not change.  
The Doppler shift picture explained in the main text works well.

\end{document}